\documentclass{aa}
\usepackage[varg]{txfonts}
\usepackage{graphicx}
\usepackage{natbib}
\usepackage{appendix}
\usepackage{listings}
\usepackage{amsmath}
\usepackage{mathtools}
\usepackage{enumitem} 
\usepackage{hyperref}

\hypersetup{
     colorlinks   = true,
     allcolors    = blue
}
\begin{document}

\titlerunning{The emergence of the \texorpdfstring{$M_D - \dot{M}_*$}{MD-M*} correlation in the MHD wind-driven scenario}
\title{The emergence of the \texorpdfstring{$M_D - \dot{M}_*$}{MD-M*} correlation in the MHD wind scenario
  } 

\author{Luigi Zallio\inst{1,2}
  \and Giovanni Rosotti\inst{1,3} 
  \and Benoît Tabone\inst{4}
  \and Leonardo Testi\inst{5}
  \and Giuseppe Lodato\inst{1}
  \and Alice Somigliana\inst{2,6}}

\offprints{L. Zallio, \email{\href{mailto:luigi.zallio@unimi.it}{luigi.zallio@unimi.it}}}

\institute{Dipartimento di Fisica ‘Aldo Pontremoli’, Università degli Studi di Milano, via G. Celoria 16, I-20133 Milano, Italy.
  \and European Southern Observatory, Karl-Schwarzschild-Strasse 2, D-85748 Garching bei München, Germany.
\and Leiden Observatory, Leiden University, P.O. Box 9513, 2300 RA Leiden, the Netherlands.
  \and Université Paris-Saclay, CNRS, Institut d’Astrophysique Spatiale, 91405 Orsay, France.
  \and Dipartimento di Fisica e Astronomia, Università di Bologna, Via Gobetti 93/2, 40122 Bologna, Italy.
  \and Fakultat für Physik, Ludwig-Maximilians-Universität München, Scheinersts. 1, 81679 München, Germany.}

\date{Received ... / Accepted ...}

\abstract{There is still much uncertainty around the mechanism that rules the accretion of proto-planetary disks. In the last years, Magnetohydrondynamic (MHD) wind-driven accretion has been proposed as a valid alternative to the more conventional viscous accretion. In particular, winds have been shown to reproduce the observed correlation between the mass of the disk $M_D$ and the mass accretion rate onto the central star $\dot{M}_*$, but this has been done only for specific conditions. It is not clear whether this implies fine tuning or if it is a general result.}{We investigate under which conditions the observed correlation between the mass of the disk $M_D$ and the mass accretion rate onto the central star $\dot{M}_*$ can be obtained.}{We present mainly analytical calculations, supported by Monte-Carlo simulations. We also perform a comparison with the observed data to test our predictions.}{We find that, in the absence of a correlation between the initial mass $M_0$ and the initial accretion timescale $t_{\mathrm{acc,0}}$, the slope of the $M_D-\dot{M}_*$ correlation depends on the value of the spread of the initial conditions of masses and lifetimes of disks. Then, we clarify the conditions under which a disk population can be fitted with a single power-law. Moreover, we derive an analytical expression for the spread of $\log \bigl(M_D/\dot{M}_*\bigr)$ valid when the spread of 
$t_{\mathrm{acc}}$ is taken to be constant. In the presence of a correlation between $M_0$ and $t_{\mathrm{acc,0}}$, we derive an analytical expression for the slope of the $M_D-\dot{M}_*$ correlation in the initial conditions of disks and at late times. In this new scenario, we clarify under which conditions the disk population can be fitted by a single power-law, and we provide empirical constraints on the parameters ruling the evolution of disks in our models.}{We conclude that MHD winds can predict the observed values of the slope and the spread of the $M_D-\dot{M}_*$ correlation under a broad range of initial conditions. This is a fundamental expansion of previous works on the MHD paradigm, exploring the establishment of this fundamental correlation beyond specific initial conditions.}

\maketitle

\section{Introduction}

Proto-planetary disks are the birth environment of planets. In the last decades, significant progress has been made in the field of sub-mm radio-astronomy, in particular thanks to the development of the Atacama Large Millimeter/Sub-millimeter Array (ALMA) observatory. Thanks to a transformational improvement in the capabilities of the new generation of telescopes, we now know that disks are sub-structured, showing gaps, spirals, kinks, and many more features (e.g., \citealt{Andrews2020}).

In addition to high resolution observations, the study of disk demographics also advanced significantly. In the last years, it was possible to collect global disk properties such as sub-mm fluxes (e.g., \citealt{Mann_2014}, \citealt{Ansdell_2016}, \citealt{Barenfeld_2016}, \citealt{Pascucci_2016},  \citealt{Ansdell_2017}, \citealt{Cox_2017}, \citealt{Eisner_2018}, \citealt{Cazzoletti_2019}, \citealt{Cieza_2019}, \citealt{Ansdell_2020}), disk radii (e.g., \citealt{Barenfeld_2017}, \citealt{Ansdell_2018}, \citealt{Sanchis2021}), and mass accretion rates onto the central stars (e.g., \citealt{Manara_2015}, \citealt{Manara_2017}, \citealt{Alcala_2017}, \citealt{Manara_2020}). This astonishing amount of new data allowed to start investigating the statistical distributions of disk properties, which make it possible to constrain evolutionary models (e.g., \citealt{Manara_2023} for a review). However, despite all the progress that has been made, we are still far from a paradigm that can completely describe the formation and evolution of disks (e.g., \citealt{Morbidelli_2016}).

Understanding how proto-planetary disks accrete material onto the central star and evolve is of cardinal importance for getting a standard model of planet formation and evolution. Two main physical scenarios have been proposed to explain accretion onto the central star: "viscous" accretion (e.g., \citealt{Shakura_Sunyaev_1973}, \citealt{Lynden-Bell_1974}), which redistributes the angular momentum within the disk (e.g., \citealt{Pringle1981}, \citealt{frank_king_raine_2002}), and "Magnetohydrodynamic wind-driven" (MHD wind-driven) accretion, where a vertical magnetic field launches a wind that extracts angular momentum from the disk (e.g., \citealt{Blandford_1982}, \citealt{Ferreira_1997}, \citealt{Lesur_2020}). Despite all the work put in developing these paradigms, there is still significant uncertainty on which is the best candidate in describing the evolution of proto-planetary disks, as their key predictions (e.g., the different time evolution of disk radii) are difficult to observe on a statistically-significant sample of observations.

In this context, observations have found a correlation between the mass of the proto-planetary disk and the mass accretion rate onto the central star (e.g., \citealt{Manara_2016}; from now on, the $M_D-\dot{M}_*$ correlation). In the viscous paradigm, the $M_D-\dot{M}_*$ correlation naturally arises with a slope close to unity, in agreement with observations (e.g., \citealt{Lodato2017}, \citealt{Rosotti2017}); on the other hand, it is still not clear which conditions can produce it in the MHD scenario. In particular, it is uncertain whether it would be a general feature of wind-driven populations, or specific initial conditions would be needed.

Recently, 1D disk evolution models have been proposed to describe the effect of MHD disk-winds on the long term evolution of disks (\citealt{Suzuki_2016}, \citealt{Bai_2016}, \citealt{Tabone_2022}). In particular, \cite{Tabone_2022} developed analytical solutions of global disk quantities (i.e., the surface density, the characteristic disk radius, the mass of the disk, and the mass accretion rate) for the MHD wind-driven scenario, and offered some insights about retrieving the correlation. In \cite{Tabone2022b}, they showed that for a particular set of initial parameters, the MHD model can reproduce the observed correlation in the Lupus region, the spread around this trend, and the decline of the disk fraction. In particular, they qualitatively show that a linear relationship can be obtained assuming no correlation between disk mass and accretion timescale $t_{\mathrm{acc},0}$. 
Starting from the solutions of \cite{Tabone_2022}, in this work we conduct an extensive analysis to discuss the universality of this result and whether it implies fine tuning.

The scope of the present work is to go beyond the work of \cite{Tabone_2022} to quantify under which initial conditions it is possible to reproduce the observed $M_D-\dot{M}_*$ correlation. In Sect. \ref{sec:no_corr}, we show the condition under which it is possible to observe the $M_D - \dot{M}_*$ correlation under the assumption that the initial accretion timescale $t_{\mathrm{acc},0}$\footnote{From \cite{Tabone_2022}: the initial accretion time-scale is a generalization of the viscous time scale and corresponds to the time that would be required to accrete a fluid particle located initially at $r_c(t=0)/2$ to the inner region of the disk with an accretion velocity equal to its initial value.} is not correlated with $M_0$. After that, we analytically derive the slope of the $M_D-\dot{M}_*$ correlation and its spread as function of the spread of the disk age $t$ and $t_{\mathrm{acc},0}$. In Sect. \ref{sec:corr}, we relax the assumption of no correlation between $M_0$ and $t_{\mathrm{acc},0}$: we introduce a correlation in the form $M_0 \propto t_{\mathrm{acc},0}^{\phi}$, and we analytically derive the slope of the $M_D - \dot{M}_*$ correlation in the initial lifetime conditions of the disks and at late times. Then, by adding a spread in $M_0$, we show that we can retrieve the observed value of the slope under a broad range of conditions. In Sect. \ref{sec:data_section}, we show that the spread derived in Sect. \ref{sec:no_corr} agrees with the observed spread of $\log \bigl(M_D / \dot{M}_*\bigr)$.
Finally, we present and investigate the $\phi - \sigma_{M_0}$ degeneracy. In Sect. \ref{sec:conclusions} we summarize our conclusions.

\section{The \texorpdfstring{$M_D-\dot{M}_*$}{MD-M*} correlation in the absence of a \texorpdfstring{$M_0-t_{\mathrm{acc,0}}$}{M0-tacc,0} correlation}
\label{sec:no_corr}

\cite{Tabone_2022} find that a proto-planetary disk evolving under the effect of MHD wind torque is governed by the master equation
\begin{equation}
\begin{split}
    \frac{\partial \Sigma}{\partial t} & = \frac{3}{r} \frac{\partial}{\partial r} \biggl\{ \frac{1}{r \Omega} \frac{\partial}{\partial r} \bigl( r^2 \alpha_{\mathrm{SS}} \Sigma c_{s}^2  \bigr)\biggr\} + \frac{3}{2r} \frac{\partial}{\partial r} \biggl\{ \frac{\alpha_{\mathrm{DW}} \Sigma c_{s}^2}{\Omega}  \biggr\} \\
    & - \frac{3 \alpha_{\mathrm{DW}} \Sigma c_{s}^2}{4 (\lambda-1)r^2 \Omega}, 
\end{split}
\label{eq:master_equation}
\end{equation}
where $\alpha_{\mathrm{SS}}$ is the Shakura-Sunyaev $\alpha$-parameter, while $\alpha_{\mathrm{DW}}$ is its analog for a MHD wind, and $\lambda$ is the magnetic lever arm. In particular, $\alpha_{DW}$ describes the amount of angular momentum extracted by the wind, while $\lambda$ determines the mass-loss rate of the wind.

When the viscous torque is neglected (i.e., when MHD winds alone describe the disk evolution and dispersal), assuming $\alpha_{\mathrm{DW}}$ and $\lambda$ to be constant across the disk, and $\lambda$ to be constant in time, it is possible to solve analytically the master equation, obtaining the surface density profile
\begin{equation}
    \Sigma(r,t) = \Sigma_{\mathrm{c}} (t) (r/r_{\mathrm{c}})^{-1+\xi} \mathrm{e}^{-r/r_{\mathrm{c}}},
    \label{eq:density_profile}
\end{equation}
where $\xi = 1/ [2(\lambda-1)]$, and $r_{\mathrm{c}}$ is the disk characteristic radius.
\cite{Tabone_2022} present solutions where $\alpha_{DW}$ increases as the disk mass decreases, namely as
\begin{equation}
    \alpha_{DW} (t) = \alpha_{DW}(0) (\Sigma_c(t)/\Sigma_c(0))^{-\omega},
\end{equation}
so that $\alpha_{DW}$ accelerates the evolution of the disk.
Here, $\omega$ is a phenomenological parameter between 0 and 1 that describes the unknown dissipation of the magnetic field. In particular, $\omega$ describes the tendency of the magnetization to increase as the disk mass decreases: for example, the case $\omega=1$ mimics a disk where the magnetic field strength is constant over time, while the case $\omega=0$ represents a disk with constant magnetization, so that it is dispersed in the same time-scale of the mass dissipation.
In this set of solutions, the disk mass evolves as
\begin{equation}
\label{eq:M_D}
M_D(t) = M_0 \> \Biggl(1-\frac{\omega t}{2 t_{\mathrm{acc},0}}\Biggr)^{1/\omega},
\end{equation}
and the mass accretion rate $\dot{M}_*$ as
\begin{equation}
\label{eq:M_dot_star}
\dot{M}_*(t)= \frac{M_0}{2t_{\mathrm{acc},0}(1+f_{M,0})} \Biggl(1-\frac{\omega t}{2 t_{\mathrm{acc},0}} \Biggr)^{1/ \omega \> -\>1},
\end{equation}
where $f_{M,0}$ is the initial mass ejection-to-accretion ratio
\begin{equation}
    f_{M,0} = (r_{\mathrm{c}}/r_{\mathrm{in}})^{\xi}-1.
\end{equation}
In the end, $M_D(t)$ and $\dot{M}_*(t)$ are controlled by the four independent parameters $M_0$, $f_{M,0}$, $t_{\mathrm{acc,0}}$ and $\omega$. In this work, if not otherwise specified we set $f_{M,0}=2.2$ and $\omega = 0.8$.

In this paper, we test the emergence of the $M_D-\dot{M}_*$ correlation. In principle, the accretion onto the central star $\dot{M}_*$ happens in the innermost region of the disk, where this latter is expected to be turbulent (e.g., \citealt{Najita_1996}, \citealt{Carr_2004}, \citealt{Hartmann_2004}, \citealt{Najita_2009}, \citealt{Ilee_2014}), while in the outer region we assumed that the disk is non-turbulent. However, from the modeling point of view, a small turbulent inner disk will always simply re-adjust to the mass accretion rate to which it is fed by the outer wind-driven disk, since locally the viscous timescale is smaller than the disk age. In particular, the accretion rate onto the central star will always quickly adjust to the accretion rate at the inner edge of the non-turbulent region. Hence, we do not explicitly model an inner turbulent region.

\subsection{The emergence of the \texorpdfstring{$M_D-\dot{M}_*$}{MD-M*} correlation}
Studying populations of disks of similar age, it is of great interest to consider the concept of \textit{isochrones} (introduced for the first time in \citealt{Lodato2017}), defined as the location in the $M_D-\dot{M}_*$ plane of the sample of disks that have the same initial mass $M_0$, same age $t$, and different accretion timescales $t_{\mathrm{acc},0}$.

\begin{figure}[t]
    \centering
    \includegraphics[width=\linewidth]{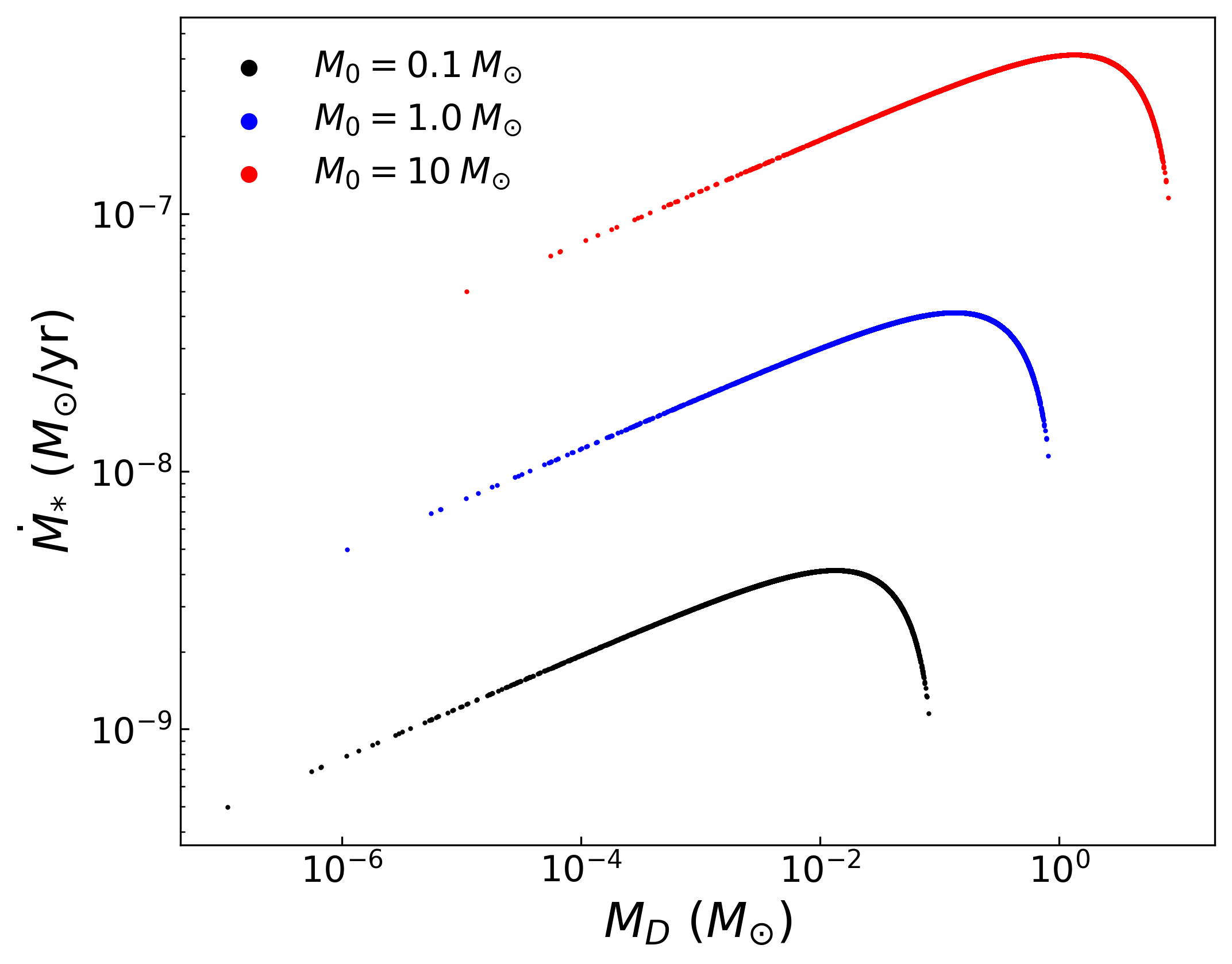}
    \caption{Isochrones of a population of disks with $t=5$ Myrs for three values of $M_0$ in the $M_D-\dot{M}_*$ plane. The colored dots are Monte-Carlo simulations of $10^5$ proto-planetary disks evolving following Eqs. \eqref{eq:M_D}, \eqref{eq:M_dot_star} with $\omega=0.8$, where $t_{\mathrm{acc},0}$ follows a Lognormal distribution centered on the natural logarithm of $1$ Myrs with a spread of 0.26 dex, while $M_0 =0.1$, $1.0$, $10 \> M_{\odot}$.}
    \label{fig:Isochrones}
\end{figure}
In Fig. \ref{fig:Isochrones} we show the  shape of the isochrones of a population of disks. Every dot is a proto-planetary disk, for which the value of $M_D$ and $\dot{M}_*$ are evaluated using Eqs. \eqref{eq:M_D}, \eqref{eq:M_dot_star}: $t_{\mathrm{acc},0}$ follows a Lognormal distribution centered\footnote{We recall that, for a Lognormal distribution, the mean value is different from the central value of the distribution. Therefore, for this work, we always indicate the median values of our Lognormal distributions and the spreads of their underlying Normal distributions.} on the natural logarithm of $1$ Myrs with a spread of 0.26 dex, while $M_0=0.1$, $1.0$, $10 \>M_{\odot}$. We recover the analytical shape of the isochrone provided in \cite{Tabone_2022}. As shown, these curves cannot intrinsically represent a disk population that follows the observed $M_D-\dot{M}_*$ correlation: the "boomerang-shaped" isochrones cannot be represented by a power-law in the form $\dot{M}_* \propto M_D^{1-\omega}$. These peculiar curves show that disks with high mass and low accretion rate are the ones with the longest $t_{\mathrm{acc,0}}$, while disks with the lowest mass and low accretion rates are the ones with shortest $t_{\mathrm{acc,0}}$.

\begin{figure*}
    \sidecaption
    \centering
    \includegraphics[width=0.6\linewidth]{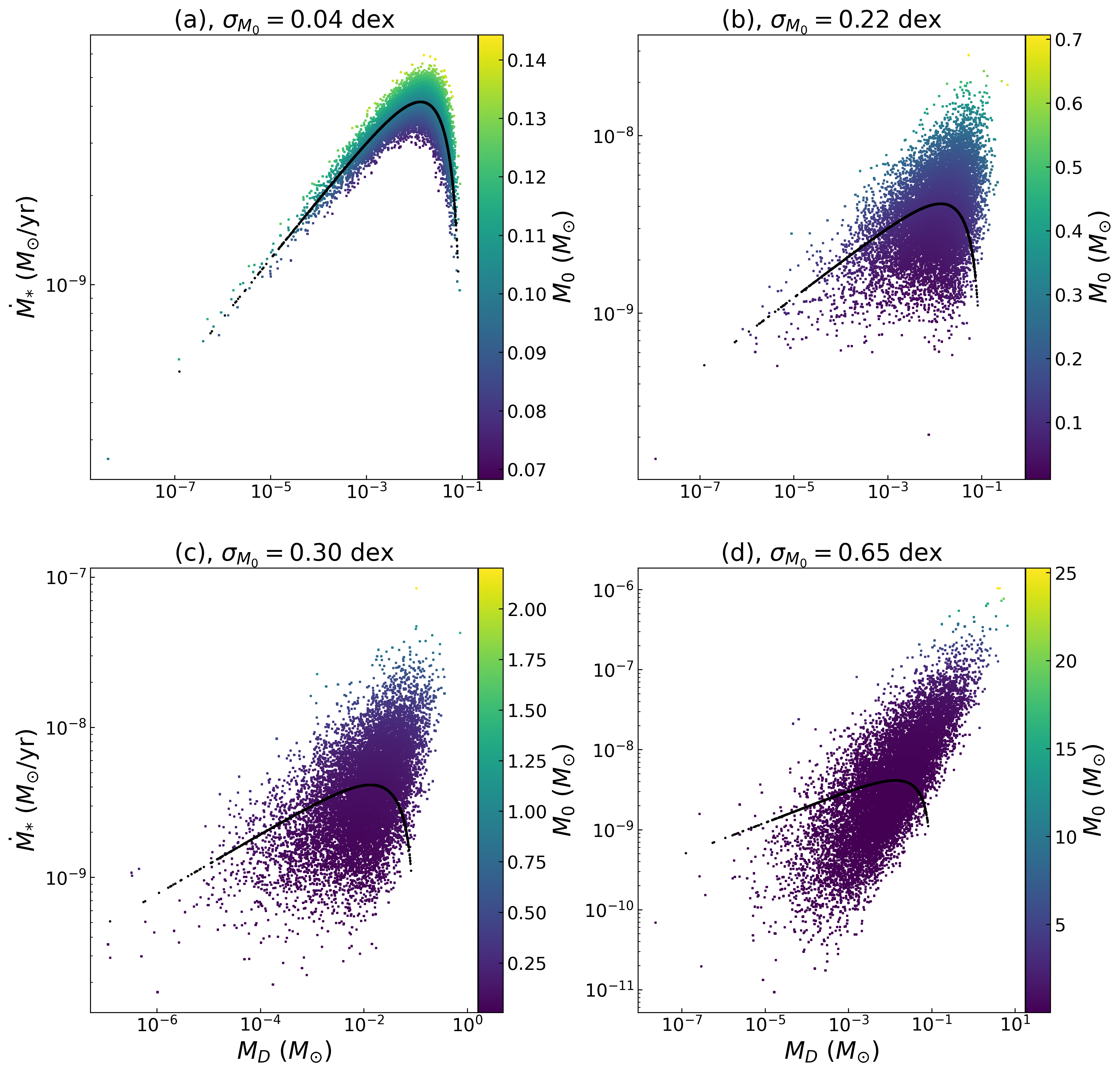}
    \caption{The $M_D-\dot{M}_*$ plane for different values of $\sigma_{M_0}$. The colored dots in every plot are Monte-Carlo simulations of $10^5$ proto-planetary disks with $\omega=0.8$ and $t=5$ Myrs, evolving following Eqs. \eqref{eq:M_D}, \eqref{eq:M_dot_star}, where $M_0$ follows a Lognormal distribution centered on the natural logarithm of $0.1\>M_{\odot}$ with a spread of (a) 0.04 dex, (b) 0.22 dex, (c) 0.30 dex, (d) 0.65 dex. The black dots show the "boomerang" shape of the isochrones for a disk population with the same parameters as in the other disk populations, but with $\sigma_{M_0}=0$. For the four plots, $t_{\mathrm{acc},0}$ follows a Lognormal distribution centered on the natural logarithm of $1$ Myrs, with a spread of 0.26 dex.} 
    \label{fig:multiplot}
\end{figure*}

To visualize the $M_D -\dot{M}_*$ correlation we set $M_0$ to follow a Lognormal distribution rather than taking it constant. We show in Fig. \ref{fig:multiplot} the results for different values of its spread. We find that the "boomerang" shape of isochrones, as defined above, is visible in the $M_D-\dot{M}_*$ plane only if the spread of the initial accretion timescale is large enough. Roughly, we can say it tends to be visible for $\sigma_{t_{\mathrm{acc},0}} > \sigma_{M_0}$, where $t_{\mathrm{acc},0}$ is the initial accretion timescale of the disk, while in the opposite case the points follow a power-law behavior.

We underline that the observed $M_D - \dot{M}_*$ correlation (\citealt{Manara_2016}) is a power-law with slope close to unity. In order to sistematically determine the values of $\sigma_{M_0}$ and $\sigma_{t_{\mathrm{acc,0}}}$ that lead to a correlation between $M_D$ and $\dot{M}_*$, we use the coefficient of determination $R^2$. Combining a visual inspection of the $M_D - \dot{M}_*$ plane with the evaluation of $R^2$ from the best power-law fit for each disk population, we derived a conservative ``rule of thumb'': $R^2 \gtrsim 0.5$. If $R^2 \gtrsim 0.5$, the disk population can be safely fitted in the $M_D-\dot{M}_*$ with a single power-law. We stress that this general rule is only informative of the goodness of fitting a disk population in the $M_D-\dot{M}_*$ plane with a single power-law, and that later in this section we investigate under which conditions we can derive a linear correlation between $M_D$ and $\dot{M}_*$. Moreover, we underline that our criterion is in agreement with the correlation coefficient found in \cite{Manara_2016} of $r = (0.56\pm 0.12)$, which changes to $r=(0.7 \pm 0.1)$ when the upper limits are considered in the analysis.

\begin{figure*}
    \sidecaption
    \centering
    \includegraphics[width=0.6\linewidth]{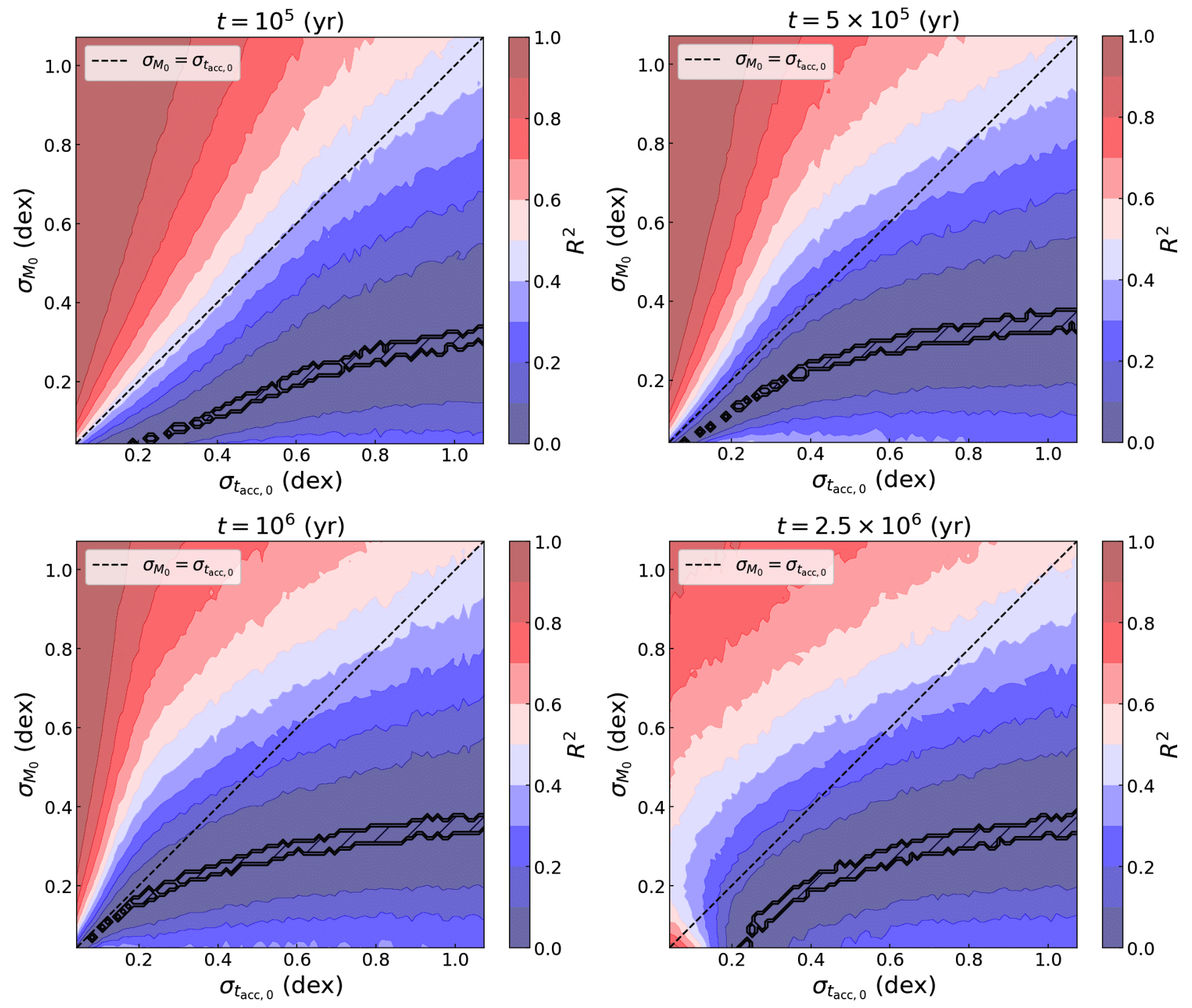}
    \caption{The $\sigma_{M_0}-\sigma_{t_{\mathrm{acc,0}}}$ plane with the values of $R^2$ for different values of age $t$. The colored dots in every plot are Monte-Carlo simulations of $10^5$ proto-planetary disks with $\omega=0.8$, evolving following Eqs. \eqref{eq:M_D}, \eqref{eq:M_dot_star}, where $M_0$ and $t_{\mathrm{acc,0}}$ follow a Lognormal distribution centered on the natural logarithm of $0.1\>M_{\odot}$ and $1$ Myrs, respectively. The hatched contours represent the regions where $R^2 = 0$, while the dashed line represents the points where $\sigma_{M_0} = \sigma_{t_{\mathrm{acc,0}}}$.} 
    \label{fig:sigmaM0-sigmatacc0_omega=0.8}
\end{figure*}

In Fig. \ref{fig:sigmaM0-sigmatacc0_omega=0.8} we show the value of $R^2$ (the coefficient of determination) measured on a large grid of models.
In these plots we set $M_0$ and $t_{\mathrm{acc,0}}$ to follow a Lognormal distribution centered on the natural logarithm of $0.1$ $M_{\odot}$ and $1$ Myrs, respectively, and we chose $\omega=0.8$. The hatched contours represent the regions where $R^2 = 0$, and the dashed line represents the points where $\sigma_{M_0} = \sigma_{t_{\mathrm{acc,0}}}$.
From Fig. \ref{fig:sigmaM0-sigmatacc0_omega=0.8}, we can see the regions of the $\sigma_{M_0}-\sigma_{t_{\mathrm{acc,0}}}$ plane that allow for a small or high value of $R^2$. Starting from these plots, we can extract different combinations of values for $\sigma_{M_0}$ and $\sigma_{t_{\mathrm{acc,0}}}$, and visualize which of them, at different ages and for different values of $\omega$, allow to obtain a power-law correlation in the $M_D-\dot{M}_*$ plane. We chose to span from very young ($t=10^5$ yr) to evolved disk populations ($t=2.5 \times 10^6$ yr) to fully explore the parameter space. 

\cite{Tabone2022b} have shown that there is no requirement to fine-tune $\omega$ to reproduce the observed accretion rates and disk masses, although explaining the rapidity of disk dispersal disfavors values of $\omega \sim 0$; in particular, if $0.5 \lesssim \omega <1$ these latter experience a steeper drop before being dispersed and in this way reproduce the observed distribution. Hence, we chose $\omega=0.8$, which is bracketed by the $\omega=0.5$ and $\omega=1$ solutions shown by \cite{Tabone2022b}.
For completeness, in Appendix \ref{appendix:sigma_M0-sigma_tacc0}, we show the same plots as in Fig. \ref{fig:sigmaM0-sigmatacc0_omega=0.8}, but for the extreme cases of $\omega=0.2$ and $\omega=1$, highlighting how the results of this work hold also for different values of $\omega$.

From Fig. \ref{fig:sigmaM0-sigmatacc0_omega=0.8}, we note that our criterion $R^2 \gtrsim 0.5$ is very roughly described by
\begin{equation}
    \frac{\sigma_{M_0}}{\sigma_{t_{\mathrm{acc,0}}}}>1.
    \label{eq:sigmaM0>sigmatacc0}
\end{equation}
In general, if Eq. \eqref{eq:sigmaM0>sigmatacc0} is satisfied, the population of disks that we consider can be described, in first approximation, as having constant $t_{\mathrm{acc},0}$. For such a population, the slope of the $M_D-\dot{M}_*$ correlation is (for the analytical calculations, refer to Appendix \ref{appendix:tacc0_const}) $\gamma = 1$. 

\begin{figure}[t]
  \includegraphics[width=\linewidth]{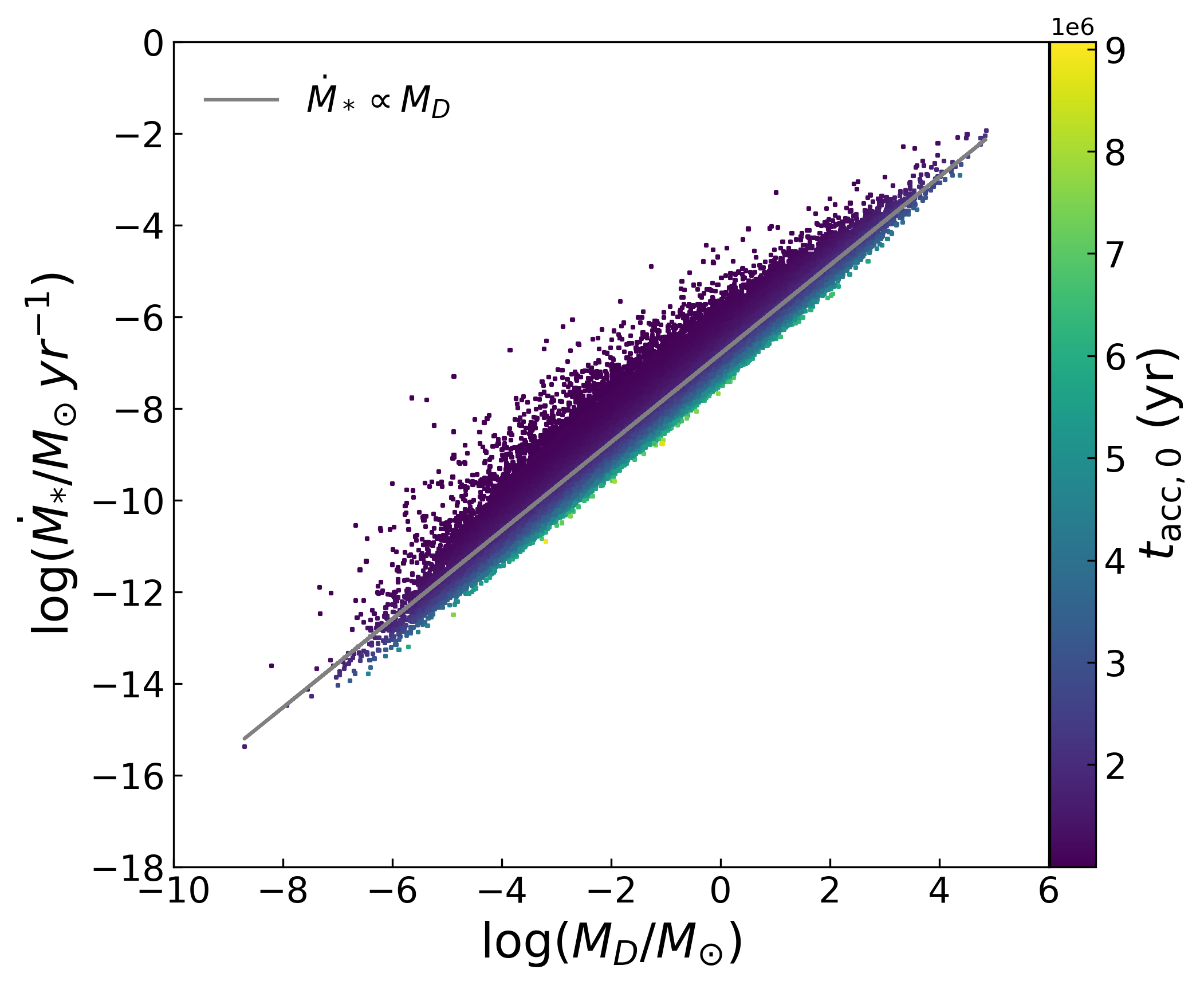}
  \caption{The $M_D - \dot{M}_*$ correlation with $\omega = 0.8$ and $t=2$ Myrs. The colored dots are Monte-Carlo simulations of $10^5$ proto-planetary disks evolving following Eqs. \eqref{eq:M_D}, \eqref{eq:M_dot_star} where $M_0$ and $t_{\mathrm{acc},0}$ follow a Lognormal distribution centered, respectively, on $0.1 \> M_{\odot}$ with a spread of 1.5 dex and on $2$ Myrs with a spread of 0.15 dex. The gray line represents the best fit of a single power-law, for which $\gamma =1$.}
  \label{fig:No_Corr}
\end{figure}

In Fig. \ref{fig:No_Corr} we show the $M_D - \dot{M}_*$ correlation in the regime described by Eq. \eqref{eq:sigmaM0>sigmatacc0}. Here, $M_0$ and $t_{\mathrm{acc},0}$ follow a Lognormal distribution centered on the natural logarithm of $0.1 \> M_{\odot}$ with a spread of 1.5 dex and on the natural logarithm of $2$ Myrs with a spread of 0.15 dex, respectively. The gray line is the fit of a single power-law, which results in $\gamma =1$. Observing Fig. \ref{fig:No_Corr}, it is important to remark that, if $\sigma_{t_{\mathrm{acc},0}}>\sigma_{M_0}$, we would not be observing a linear correlation between $M_D$ and $\dot{M}_*$: we would instead see the "boomerang" isochrone shape shown in Fig. \ref{fig:Isochrones}, which depends likewise on the value of $\omega$.

We stress that Eq. \eqref{eq:sigmaM0>sigmatacc0} is only an approximate criterion. The exact criterion under which the $M_D-\dot{M}_*$ correlation is a power-law with a slope of unity is more complex and depending on the age and on the value of $\omega$. In particular, we find that for low values of $\omega$ (see Appendix \ref{appendix:sigma_M0-sigma_tacc0}) the correlation is harder to get, in particular at late time. For detailed studies, we thus recommend using Figs. \ref{fig:sigmaM0-sigmatacc0_omega=0.8}, \ref{fig:sigmaM0-sigmatacc0_omega=0.2}, \ref{fig:sigmaM0-sigmatacc0_omega=1.0} to identify the regions of the parameters space that give a correlation instead of Eq. \ref{eq:sigmaM0>sigmatacc0}.

We can summarize the results obtained as follows: 
\begin{enumerate}
    \item In the case in which there is no spread in $M_0$, a "boomerang-shaped" isochrone appears in the $M_D-\dot{M}_*$ plane.
    \item In the limit $\sigma_{M_0}>>\sigma_{t_{\mathrm{acc,0}}}$, the slope of the $M_D-\dot{M}_*$ relation is exactly 1, a result that can be obtained also analytically.
    \item In the general case, we find that the correlation is recovered when the spread in initial disk mass is typically larger than the spread in initial accretion timescale. This rule of thumb can be refined using Figs. \ref{fig:sigmaM0-sigmatacc0_omega=0.8}, \ref{fig:sigmaM0-sigmatacc0_omega=0.2}, \ref{fig:sigmaM0-sigmatacc0_omega=1.0}.
\end{enumerate}
However, observations provide us with more than a correlation, they give the spread around this correlation and the power-law index. In particular, observational results (e.g., \citealt{Manara_2016}, \citealt{Testi_2022}) consistently find values of the $M_D-\dot{M}_*$ correlation which are <1 along with a large spread. Therefore, in the next section, we study under which conditions we can retrieve the observed value of the slope of the $M_D-\dot{M}_*$ correlation and its spread.

\subsection{The slope and the spread of the \texorpdfstring{$M_D-\dot{M}_*$}{MD-M*} correlation as function of $\sigma_{M_0}$ and $\sigma_{t_{\mathrm{acc,0}}}$.}
\label{sec:slope_MD_Mdot}
\begin{figure}[t]
    \centering
    \includegraphics[width=\linewidth]{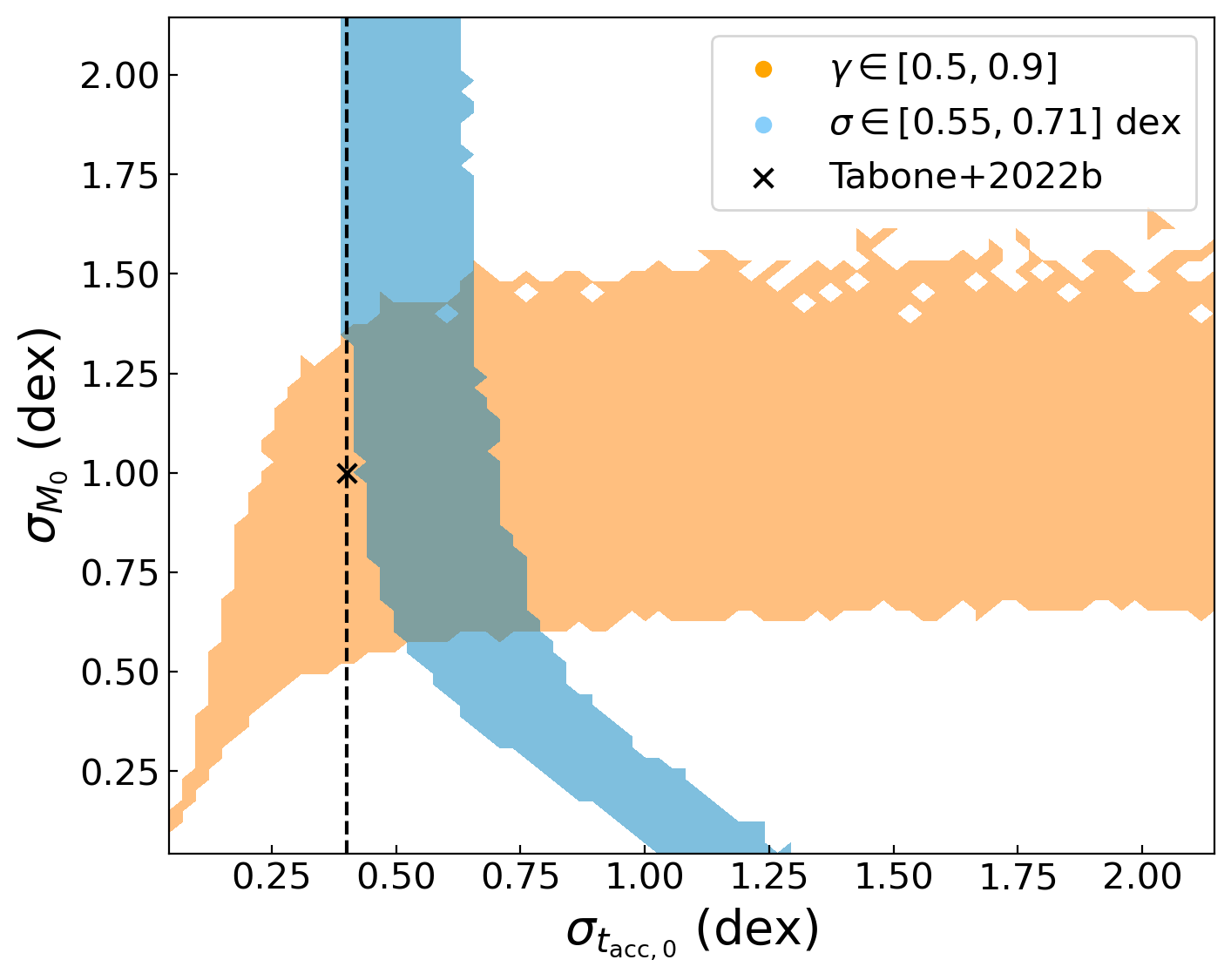}
    \caption{The $\sigma_{t_{\mathrm{acc,0}}} - \sigma_{M_0}$ plane with the values of the slope $\gamma$ and the spread of the $M_D-\dot{M}_*$ correlation $\sigma$: we put in evidence in orange the values $\gamma \in [0.5, 0.9]$, while in light blue the values $\sigma \in [0.55, 0.71]$ dex, as they confine the region of values compatible with the values found in \cite{Manara_2016}. The colored regions show the value of $\gamma$ and $\sigma$ for Monte-Carlo disk populations obtained with the corresponding specific values of $\sigma_{t_{\mathrm{acc,0}}}$ and $\sigma_{M_0}$. Every Monte-Carlo consists in $10^4$ disks ruled by Eqs. \eqref{eq:M_D}, \eqref{eq:M_dot_star}, where $\omega = 0.8$, $t=2$ Myrs, while $M_0$ and $t_{\mathrm{acc},0}$ follow a Lognormal distribution centered on the natural logarithm of $0.1\>M_{\odot}$ and $2$ Myrs, respectively. The cross shows the case investigated by \cite{Tabone2022b}, while the dashed line highlights the value $\sigma_{t_{\mathrm{acc},0}}=0.4$.}
    \label{fig:sigmaM0_sigmatacc0}
\end{figure}

In order to constrain the values of $\sigma_{M_0}$ and $\sigma_{t_{\mathrm{acc,0}}}$ that can reproduce the observed slope and spread of the $M_D-\dot{M}_*$ correlation, we perform a linear regression of the disks in the $M_D-\dot{M}_*$ plane for different combinations of $\sigma_{M_0}$ and $\sigma_{t_{\mathrm{acc,0}}}$. In Fig. \ref{fig:sigmaM0_sigmatacc0} we show the results of our analysis.
Here, the colored stripes in the $\sigma_{t_{\mathrm{acc,0}}}-\sigma_{M_0}$ plane represent the regions of the parameter space which predict values of the slope $\gamma$ and of the spread $\sigma$ of the $M_D-\dot{M}_*$ correlation in agreement with the observed values. In a similar fashion to the analysis performed in \cite{Lodato2017}, we show in orange the region $\gamma \in [0.5, 0.9]$, while in light blue the region $\sigma \in [0.55, 0.71]$ dex. In contrast with \citealt{Lodato2017} and \citealt{Tabone2022b}, we neglect the effect of accretion variability and the uncertainties on the mass estimates. It means that the populations tend to underestimate the spread in $M_D / \dot{M}_*$. In this plot, $t=2$ Myrs, $\omega=0.8$, and $M_0$ and $t_{\mathrm{acc},0}$ follow a Lognormal distribution centered on the natural logarithm of $0.1\>M_{\odot}$ and $2$ Myrs, respectively. The cross represents the work of \cite{Tabone2022b}. Fig. \ref{fig:sigmaM0_sigmatacc0} shows that it possible to retrieve the observed values of the slope and the spread of the $M_D-\dot{M}_*$ correlation for $\sigma_{M_0} \in [0.50, 1.50]$ and $\sigma_{t_{\mathrm{acc,0}}} \in [0.40, 0.75]$.

We stress that the results here obtained are valid for a fixed age $t$, a value of $\omega=0.8$, and the fixed median values $M_0=0.1\>M_{\odot}$ and $t_{\mathrm{acc,0}}=2$ Myrs. As we show in Appendix \ref{appendix:sigma_M0-sigma_tacc0}, changing $\omega$ will change the derived values of $\sigma_{M_0}$ and $\sigma_{t_{\mathrm{acc,0}}}$, but it remains true that it is possible to retrieve the observed slope and the spread of the $M_D-\dot{M}_*$ correlation for a wide range of spreads in disk mass and accretion timescale in the initial conditions. We also note that in Fig. \ref{fig:sigmaM0_sigmatacc0}, not all the populations fit the disk fraction. As discussed below and in \cite{Tabone2022b}, a typical value of $\sigma_{t_{\mathrm{acc},0}}\simeq0.4$ dex needs to be adopted to fit the observed decline of disk fraction with cluster age.

\subsection{The spread of \texorpdfstring{$\log \bigl( M_D/\dot{M}_* \bigr)$}{log(MD/M*)} as function of the spread in \texorpdfstring{$t$}{t} and \texorpdfstring{$t_{\mathrm{acc}}$}{tacc0}}
\label{sec:spread}
\cite{Testi_2022}, \cite{Somigliana2023} show that the spread of the accretion timescale $M_D / \dot{M}_*$ is a fundamental observable for disk population studies. Considering Eqs. \eqref{eq:M_D}, \eqref{eq:M_dot_star}, we can evaluate the dimensionless spread of the accretion timescale when $\dot{M}_* \propto M_D$. Using the propagation of errors (see more details in Appendix \ref{appendix:spread}) on $\log(M_D / \dot{M}_*)$, the dimensionless spread is
\begin{equation}
\begin{split}
    \sigma \bigl( \log \bigl( M_D / \dot{M}_* \bigr)\bigr) & \simeq \frac{1}{2\langle{{t_{\mathrm{acc}, 0}}}\rangle - \omega \langle{t}\rangle} \\
    & \sqrt{\bigl(2 \langle{{t_{\mathrm{acc}, 0}}}\rangle \>\sigma_{t_{\mathrm{acc}, 0}}\bigr)^2 + \bigl(\omega \langle{t}\rangle \>\sigma_{t}\bigr)^2}, 
\end{split}
\label{eq:sigma_2}
\end{equation}
where we use the notation $\langle{{t_{\mathrm{acc}, 0}}}\rangle$ and $\langle{t}\rangle$ to remark this is the median value for the population, and we have also allowed for the possibility of an age spread $\sigma_t$ in the population.
This result is equivalent to the spread of the $M_D-\dot{M}_*$ correlation for a slope mildly different than 1, but we stress that if the $M_D-\dot{M}_*$ correlation differs substantially from linearity, then Eq. \eqref{eq:sigma_2} is not valid anymore.

The fundamental assumption underlying Eq. \eqref{eq:sigma_2} is that $\sigma_{t_{\mathrm{acc}}}$ is taken to be constant in time. In reality, this value should be computed only over the population of surviving disks. It is not necessarily true that it stays constant, as, while evolving with time, a disk population loses the fastest evolving disks: as a consequence, $\sigma_{t_{\mathrm{acc}}}$ should decrease as time passes. However, as noted in the work of \cite{Somigliana2023}, the spread on the accretion timescale decreases slowly in time in the MHD wind-driven scenario (a factor of 0.1 dex within 2 Myrs). Therefore, Eq. \eqref{eq:sigma_2} can be considered valid for young star-forming regions. Moreover, we note that the spread of $\log \bigl( M_D/\dot{M}_* \bigr)$ is equal to that of the $M_D-\dot{M}_*$ correlation when the slope of the correlation is exactly 1. Therefore, in first approximation, the expression reported in Eq. \eqref{eq:sigma_2} gives us an estimate of the order of magnitude of the spread of the $M_D-\dot{M}_*$ correlation, as further discussed in Sect. \ref{sec:empirical_spread}.

\begin{figure}[t]
  \includegraphics[width=\linewidth]{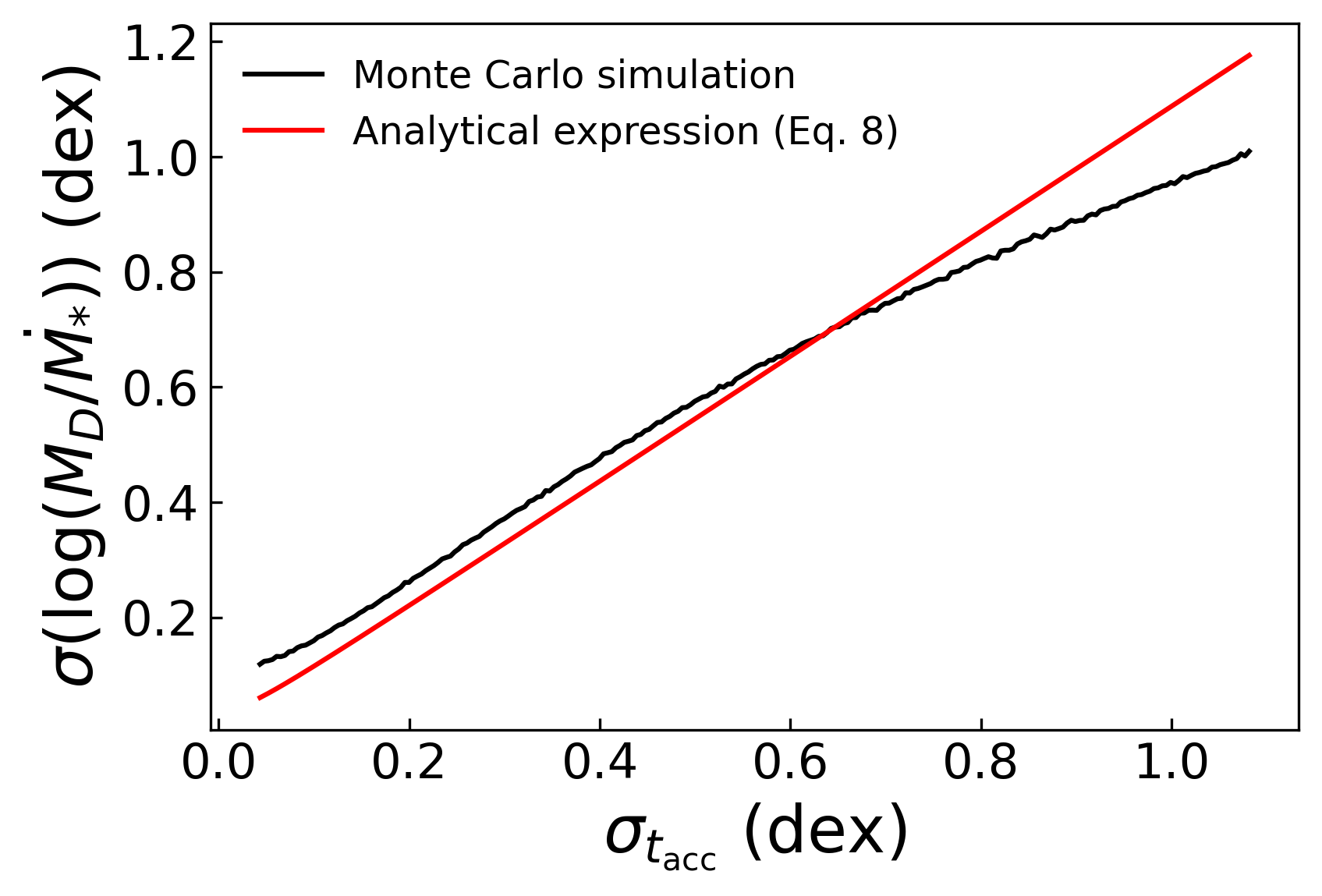}
  \caption{The spread of the $M_D - \dot{M}_*$ correlation as function of $\sigma_{t_{\mathrm{acc}}}$, with $\sigma_t=0.43$ dex. The values of $\sigma\bigl(\log\bigl(M_D/\dot{M}_*\bigr)\bigr)$ are evaluated as the standard deviation from Monte Carlo simulations of $10^5$ protoplanetary disks with $M_0$, $t_{\mathrm{acc}}$ and $t$ following a Lognormal distribution centered respectively in $0.1 \> M_{\odot}$, $5$ Myrs, $1$ Myrs.}
  \label{fig:Std_dev_sigma_t=0}
\end{figure}

In Fig. \ref{fig:Std_dev_sigma_t=0} we show the spread of $\log \bigl( M_D/\dot{M}_* \bigr)$ when $t$ follows a Lognormal distribution centered on the natural logarithm of $1$ Myrs, and $\sigma_t=0.43$ dex. The black line is computed as the standard deviation from the Monte Carlo simulations of $10^5$ disks that follow a Lognormal distribution in $t$, $t_{\mathrm{acc}}$ and $M_0$: in particular, the value of $\sigma \bigl(\log \bigl( M_D / \dot{M}_*\bigr) \bigr)$ is evaluated for $10^2$ values of $\sigma_{t_{\mathrm{acc}}}$ that span $[0, 1.1]$ dex, while $\sigma_t=0$ dex. Here, $M_0$ and $t_{\mathrm{acc}}$ follow a Lognormal distribution centered, namely, on the natural logarithm of $5$ Myrs for $t_{\mathrm{acc}}$ and on the natural logarithm of $0.1 \> M_{\odot}$ for $M_0$. We remark that the spread of $\log(M_D/\dot{M}_*)$ in independent of the spread in $M_0$ under the conditions for which Eq. \ref{eq:sigma_2} holds, which is that the slope of the $M_D-\dot{M}_*$ correlation is close to unity. In Fig. \ref{fig:sigmaM0_sigmatacc0} we see that, as we get close to $\gamma=1$, the cyan band representing the spread of the $M_D-\dot{M}_*$ correlation becomes vertical, while as we move away from those values the dependency on $\sigma_{M_0}$ appears.

As shown in \cite{Testi_2022}, \cite{Somigliana2023}, the spread on the accretion timescale is an important observable, which can be easily obtained from the data. We show a comparison with observations in Sect. \ref{sec:empirical_spread}.

Until now, we considered disk populations for which $M_0$ and $t_{\mathrm{acc,0}}$ are not correlated. First, we derived a general criterion for which we can retrieve a population of disks in the $M_D-\dot{M}_*$ plane that can be fitted with a single power-law. Secondly, we studied under which initial conditions it is possible to reproduce the observed slope and spread of the $M_D-\dot{M}_*$ correlation. Finally, we provided an approximated analytical expression for the spread on the accretion timescale $\log \bigl(M_D/\dot{M}_*\bigr)$. In the following, we study the emergence of the $M_D-\dot{M}_*$ correlation when $M_0$ is correlated to $t_{\mathrm{acc,0}}$.

\section{The \texorpdfstring{$M_D - \dot{M}_*$}{MD-M*} correlation in the presence of a \texorpdfstring{$M_0-t_{\mathrm{acc},0}$}{M0-tacc,0} correlation}
\label{sec:corr}

So far, we have considered the initial accretion timescale $t_{\mathrm{acc},0}$ to be not correlated with the initial mass of the disk $M_0$. However, this is not necessarily the case. In the following, we assume a correlation between $M_0$ and $t_{\mathrm{acc},0}$ in the form
\begin{equation}
\label{eq:M_0relt_0}
M_0 \propto t_{\mathrm{acc},0}^{\phi}.
\end{equation}
Inserting Eq. \eqref{eq:M_0relt_0} in Eqs. \eqref{eq:M_D}, \eqref{eq:M_dot_star}, it is not possible to find an analytical derivation of the slope in the form of Eq. \eqref{eq:gamma}. However, it is possible to study the analytical behavior of the slope in the initial and final stages of the evolution of the disk, starting from the definition (following \citealt{Tabone_2022}) of
\begin{equation}
\label{eq:t_disp}
t_{disp} = \frac{2 t_{\mathrm{acc},0}}{\omega}
\end{equation}
as the disk dispersal time.

\subsection{The $M_D-\dot{M}_*$ correlation at initial time}
\label{sec:initial_cond}
When taking a population of disks without spread in $M_0$, in the initial conditions (i.e. $t/t_{disp} << 1$) we find that the slope is $\gamma = 1-1/\phi$ (see Appendix \ref{appendix:t/tdisp<<1} for more details).
\begin{figure}[t]
  \includegraphics[width=\linewidth]{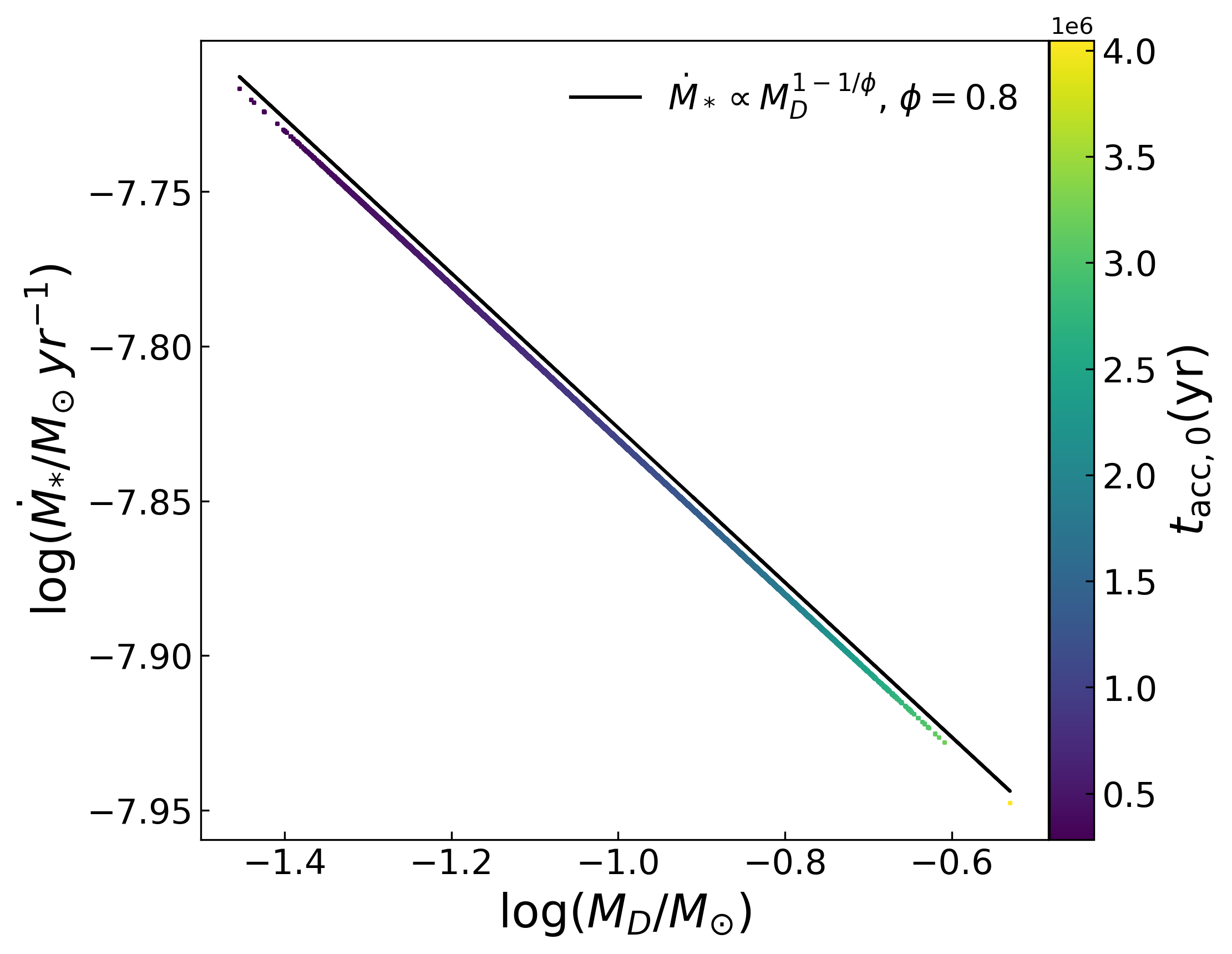}
  \caption{The slope of the $M_D - \dot{M}_*$ correlation with $\phi=0.8$, $\omega = 0.8$, and $t=0$ yr. The dots are Monte-Carlo simulations of $10^5$ proto-planetary disks ruled from Eqs. \eqref{eq:M_D}, \eqref{eq:M_dot_star} where $M_0$ follows Eq. \eqref{eq:M_0relt_0} and $t_{\mathrm{acc},0}$ follows a Lognormal distribution centered on the natural logarithm of $2$ Myrs with a spread of $0.13$ dex. The black line represents Eq. \eqref{eq:M_dot_initial}.}
  \label{fig:t_tdisp__1}
\end{figure}

In Fig. \ref{fig:t_tdisp__1} we show the $M_D - \dot{M}_*$ correlation for a disk population (dots) of $10^5$ disks ruled from Eqs. \eqref{eq:M_D}, \eqref{eq:M_dot_star}, where $M_0$ is given by Eq. \eqref{eq:M_0relt_0}. Here, $\phi=0.8$, $\omega=0.8$, $t=0$ yr, and $t_{\mathrm{acc},0}$ follows a Lognormal distribution centered on the natural logarithm of $2$ Myrs with a spread of $0.13$ dex. The black line shows the analytical $M_D-\dot{M}_*$ correlation (Eq. \eqref{eq:M_dot_initial}), which agrees remarkably well with the Monte-Carlo simulations.

\subsection{The slope close to disk dispersal}
When considering a population of disks without spread in $M_0$, close to dispersal (i.e. $t/t_{disp}\sim1$) we find that $\gamma=1-\omega$ (see Appendix \ref{appendix:t=tdisp} for more details).
This is an unexpected result, as the slope appears to reach a specific value, which is independent from the initial correlation between $M_0$ and $t_{\mathrm{acc},0}$. This is because when a disk population is close to dispersal, all the disks also share a similar $t_{\mathrm{acc},0}$ and, as a consequence, a similar $M_0$ (as they are linked to each other). Thus, when $t/t_{disp}\sim1$, the correlation between $M_0$ and $t_{\mathrm{acc},0}$ is lost, and the memory of the initial conditions is forgotten. \cite{Tabone_2022} find a similar result even without explicitly introducing a correlation. This supports the previous statement that, taking a population of disks that is about to be dispersed, it is only possible to observe the final evolutionary track of the disk (for which $\dot{M}_* \propto M_D^{1-\omega}$), thus losing any possibility of detecting the initial correlation between $M_0$ and $t_{\mathrm{acc},0}$.

\begin{figure}[t]
  \includegraphics[width=\linewidth]{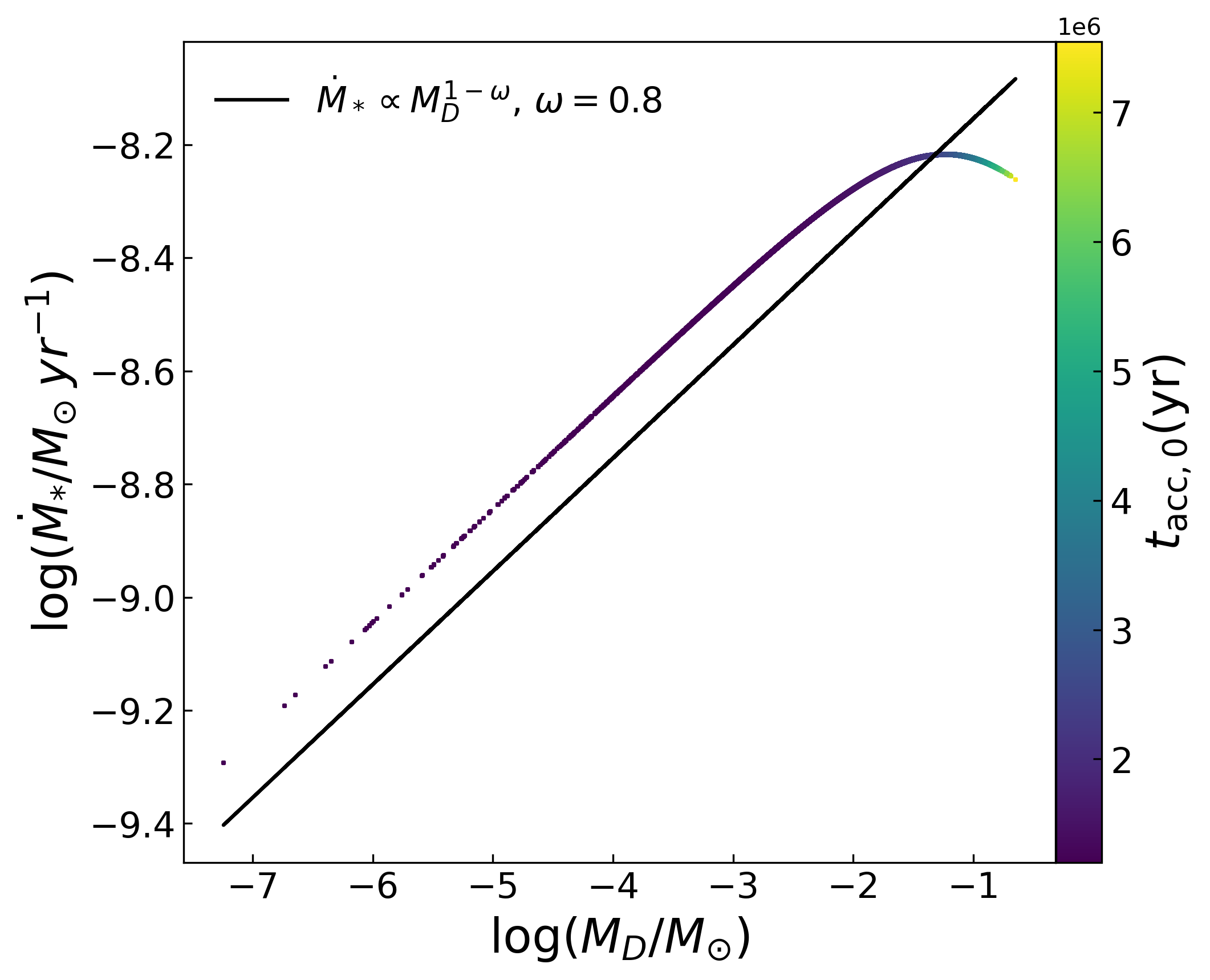}
  \caption{The slope of the $M_D - \dot{M}_*$ correlation with $\phi=0.8$, $\omega = 0.8$, and $t=3$ Myrs. The colored dots are Monte-Carlo simulations of $10^5$ protoplanetary disks ruled from Eqs. \eqref{eq:M_D}, \eqref{eq:M_dot_star} where $M_0$ follows Eq. \eqref{eq:M_0relt_0} and $t_{\mathrm{acc},0}$ follows a Lognormal distribution centered on the natural logarithm of $2$ Myrs with a spread of $0.13$ dex. The black line represents Eq. \eqref{eq:M_dot_tdisp}.}
  \label{fig:t_tdisp_sim_1}
\end{figure}

In Fig. \ref{fig:t_tdisp_sim_1} we show the $M_D - \dot{M}_*$ correlation for $\phi=0.8$, $\omega=0.8$, and $t=3$ Myrs, where the other parameters are the same as in Fig. \ref{fig:t_tdisp__1}. We see that, in this case, the $M_D - \dot{M}_*$ correlation has a bent shape, which could be better described by a double power-law. We are interested in the left part of the plot, where $M_D \lesssim 10^{-2} M_{\odot}$, i.e., when the disk lifetime is similar to the disk dispersal time. The black line shows Eq. \eqref{eq:M_dot_tdisp}: the slope of the theoretical line (black) coincides with the $M_D - \dot{M}_*$ correlation in the above cited region of the $M_D - \dot{M}_*$ plane, showing that when $t \sim t_{disp}$, the slope approaches the value $\gamma = 1-\omega$. 
The right part of Fig. \ref{fig:t_tdisp_sim_1} instead shows the disks that are still evolving, hence still near the condition $t/t_{disp}<<1$: we see that the slope here inverts its trend, in agreement with the value $1-1/\phi$ that we derived in Sect. \ref{sec:initial_cond}.

In Appendix \ref{appendix:spread_propto} we show the calculation of the spread of the $M_D-\dot{M}_*$ correlation when $M_0 \propto t_{\mathrm{acc},0}^{\phi}$. The (trivial) results are in continuity with what we have already found in Sect. \ref{sec:spread}.

\subsection{The role of \texorpdfstring{$\sigma_{M_0}$}{sigmaM0} in the \texorpdfstring{$M_D-\dot{M}_*$}{MD-M*} correlation}
\label{subsec:spread_slope}
\begin{figure}[t]
\centering
  \includegraphics[width=\linewidth]{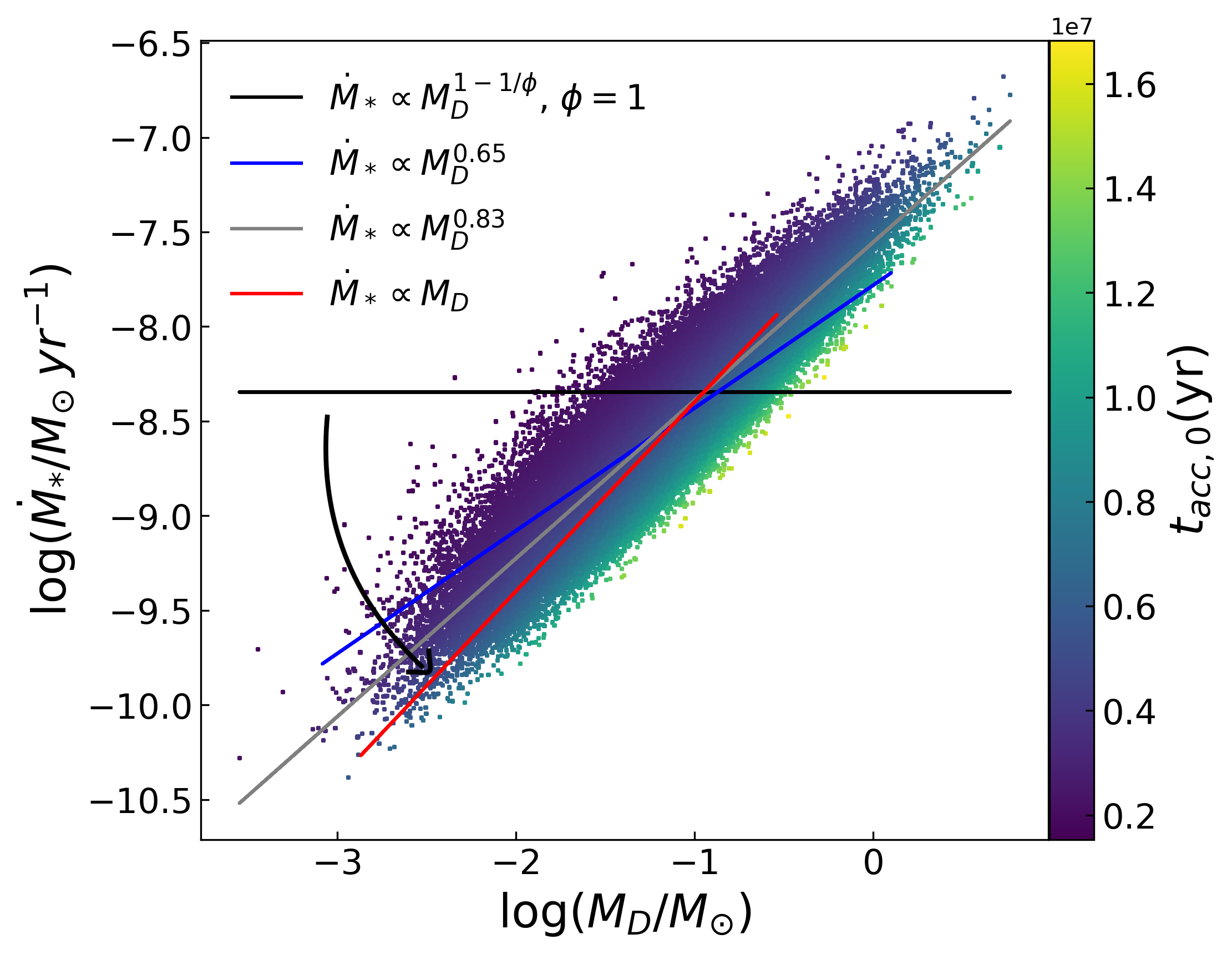}
  \caption{The $M_D - \dot{M}_*$ correlation with $\phi=1$, $\omega = 0.8$ and $t=3.5$ Myrs. The dots are Monte-Carlo simulations of $10^5$ proto-planetary disks evolving following Eqs. \eqref{eq:M_D}, \eqref{eq:M_dot_star} where $M_0$ follows Eq. \eqref{eq:M_0relt_0} and $t_{\mathrm{acc},0}$ follows a Lognormal distribution centered on the natural logarithm of $5$ Myrs with a spread of $0.13$ dex. In addition, $M_0$ is then generated following a Lognormal distribution with the values of $t_{\mathrm{acc},0}$ following Eq. \eqref{eq:M_0relt_0} with a spread of 0.43 dex. The black line represents Eq. \eqref{eq:M_dot_initial}, the gray line is the best-fitted power-law for the population, resulting in $\dot{M}_* \propto M_D^{0.83}$, while the red one shows Eq. \eqref{eq:M_dot_prop}. For illustrative purposes, we also report in blue in the plot of the best-fitted power-law for a population for which $\sigma_{M_0}=0.26$ dex, which results in $\dot{M}_*\propto M_D^{0.65}$.}
  \label{fig:Spread_M0}
\end{figure}

It is reasonable to think that, if $t_{\mathrm{acc},0}$ is generated following a Lognormal distribution with a spread, and $M_0 \propto t_{\mathrm{acc},0}^{\phi}$ is satisfied, $M_0$ should also have a spread of its own (i.e., the spread in $M_0$ reflects the fact that different stars living in the same stellar population have different masses, not only different initial timescales). To ensure that our populations have a spread in $M_0$, we initially evaluate $M_0$ with Eq. \eqref{eq:M_0relt_0}; then, $M_0$ is extracted with a Lognormal distribution, with fixed spread, centered on the previously determined values of $M_0$. In Fig. \ref{fig:Spread_M0} we show that, once the spread in $M_0$ is added, we see the appearing of the linear $M_D - \dot{M}_*$ relation. Here, we show a population of disks identical to the one in Fig. \ref{fig:t_tdisp_sim_1}, but we add a spread in $M_0$ of 0.43 dex. The addition of $\sigma_{M_0}$ hides the track shown in Fig. \ref{fig:t_tdisp_sim_1}. Moreover, the slope becomes steeper as $\sigma_{M_0}$ increases, going toward the linear behavior, which is plotted in red. As a reference, we plotted in gray the best-fitted power-law for the population, resulting in a slope of 0.83, in blue the best-fitted power-law for a population with $\sigma_{M_0}=0.26$ dex, resulting in a slope of 0.65, while in black we reported the slope obtained in Eq. \eqref{eq:gamma_alpha}.

We draw two main conclusions from this. First, it is still possible to derive the slope of the $M_D- \dot{M}_*$ correlation with a value different than 1 after the introduction of the correlation $M_0 \propto t_{\mathrm{acc},0}^{\phi}$. It is also possible to tune the value of the $M_D-\dot{M}_*$ slope changing the values of $\phi$ and $\sigma_{M_0}$: we will discuss this in Sect. \ref{sec:degeneracy}. Secondly, the addition of $\sigma_{M_0}$ deletes the canonical signature of the slope shown in Figs. \ref{fig:t_tdisp__1}, \ref{fig:t_tdisp_sim_1}.
This leads to a dramatic observational conclusion: it is not possible to directly observe the expected analytical signature of the slope because of the natural existence of spread in $M_0$.

\subsection{Can the \texorpdfstring{$M_D-\dot{M}_*$}{MD-M*} correlation still be described by a single power-law?}
It is relevant to underline that, after the introduction of a spread in $M_0$, the bent shape of the $M_D-\dot{M}_*$ correlation shown in Fig. \ref{fig:t_tdisp_sim_1} can still be recovered, but only for very low values of $\sigma_{M_0}$.\footnote{Some tests were conducted on this, but we found redundant to show the resulting plots. In particular, the bent shape shown in Fig. \ref{fig:t_tdisp_sim_1} can be recovered for $\sigma_{M_0}<0.01$ dex. In the following, we show that as a minimum amount of $\sigma_{M_0}$ is added, the $M_D - \dot{M}_*$ correlation is well described by a single power-law.} Observing Fig. \ref{fig:t_tdisp_sim_1}, it appears clear that a single power-law cannot properly fit the evolutionary track of the disks: a double power-law seems much more appropriate. Hence, we conducted a test in order to quantitatively understand when a single power-law starts to fit a disk population.

At first, we introduced a broken double power-law in the form

\begin{equation}
    \centering
    \begin{cases}
        \dot{M}_* = k_1 + a \> M_D^{-b}, \>\>\>\>\>\>\>\>\>\>\>\>\>\>\>\>\>\> M_D < x_0 \\
        \dot{M}_* = k_2 + c \> (M_D-x_0)^{d}, \>\>\>\>\>\> M_D \geq x_0
    \end{cases}
    \label{eq:double_powerlaw}
\end{equation}
where $k_2 = k_1+a\>x_0^{-b}$ is not a free parameter as the piecewise function is defined to be continuous in $M_D=x_0$. 
Then, after setting $t=3$ Myrs, we fitted both Eq. \eqref{eq:double_powerlaw} and the equation of a single power-law in the form 
\begin{equation}
    \dot{M}_* = a \> M_D^{\zeta}
    \label{eq:single_powerlaw}
\end{equation}
to the $\dot{M}_*-M_D$ plane for different values of $\sigma_{M_0}$, namely $\sigma_{M_0}\> \in \> [0, 0.022]$ dex. The other parameters for the disk population are the same as in Fig. \ref{fig:Spread_M0}. Since Eq. \eqref{eq:double_powerlaw} is defined upon six free-parameters ($x_0$, $k_1$, $a$, $b$, $c$, $d$), while Eq. \eqref{eq:single_powerlaw} upon only two ($a$, $\zeta$), we performed the fit using the Bayesian Information Criterion ($BIC$), with the python package \verb|RegscorePy.bic|. 

\begin{figure}[t]
    \centering   \includegraphics[width=\linewidth]{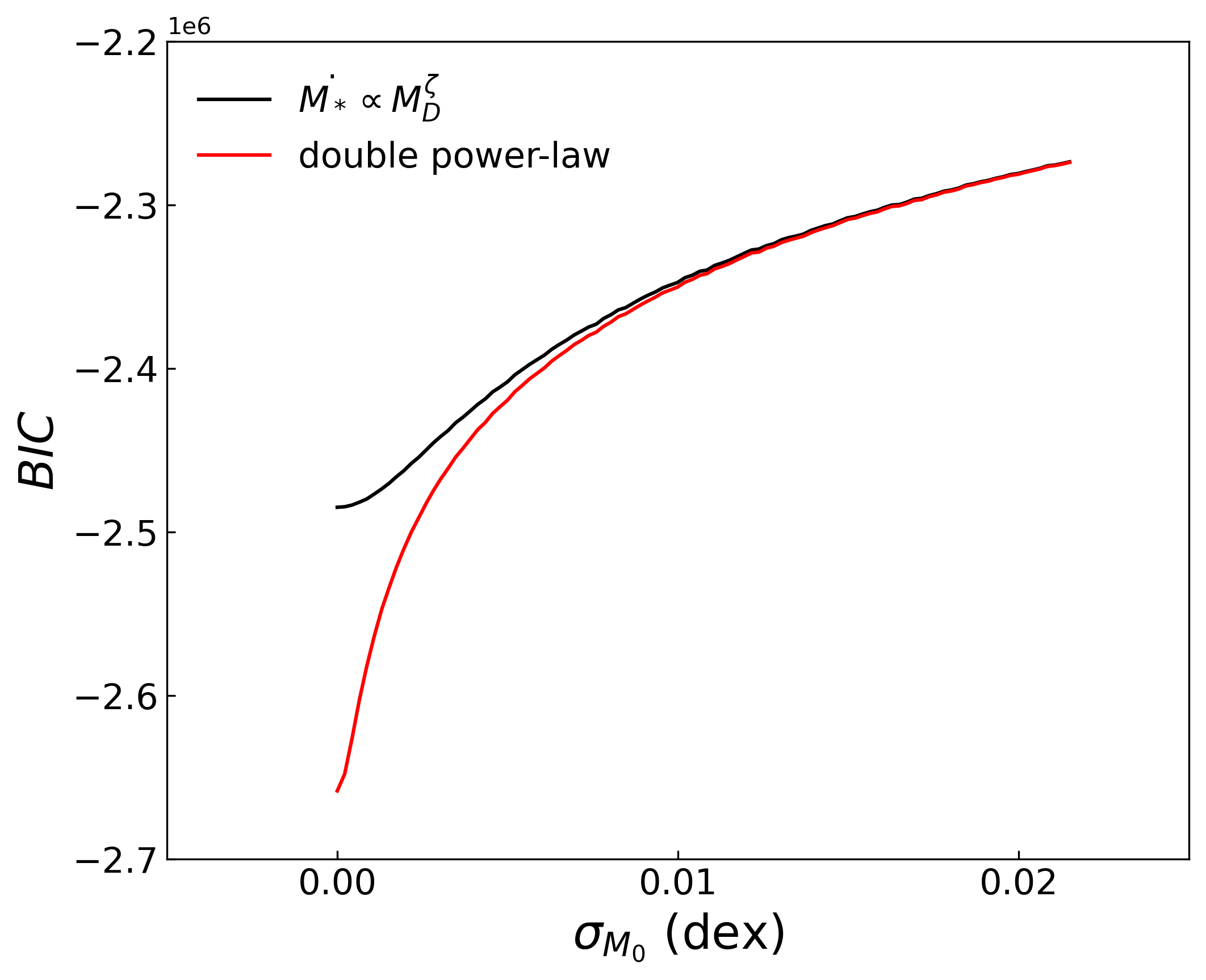}
    \caption{The Bayesian Information Criterion $BIC$ as function of $\sigma_{M_0}$ in a range $\sigma_{M_0} \> \in \> [0, 0.022]$ dex. The $BIC$ is obtained for Eq. \eqref{eq:double_powerlaw} (in red) and Eq. \eqref{eq:single_powerlaw} (in black) from the fit to the $\dot{M}_* -M_D$ plane obtained for different values of $\sigma_{M_0}$, while the simulated disk populations have the same set of parameters described in Fig. \ref{fig:t_tdisp_sim_1}.}
    \label{fig:BIC_smallspread}
\end{figure}

In Fig. \ref{fig:BIC_smallspread} we show that the $BIC$ is minimized from the double power-law in the regime $\sigma_{M_0} \in [0, 0.01]$ dex, hence showing that for very low values of $\sigma_{M_0}$, the $M_D-\dot{M}_*$ correlation is better fitted with a double power-law. However, as little spread in $M_0$ is added, the red and the black curve become indistinguishable, highlighting the fact that those disk populations can be fitted both with a single or a double power-law.\footnote{A double power-law can always be re-conducted to a single power-law: the fact that the curves perfectly superpose means that the double power-law became a single power-law.}

From this, we can conclude that when a little spread in $M_0$ is added, the $M_D-\dot{M}_*$ correlation can be described by a single instead of a double power-law. Thus, it is reasonable to use a single power-law to fit the $M_D-\dot{M}_*$ correlation.
We show the same plot of Fig. \ref{fig:BIC_smallspread} in Appendix \ref{appendix:likelihood_BIC} for $\sigma_{M_0} \> \in \> [0, 0.43]$ dex.

\section{Comparison with the data: empirical constraints on the spreads}
\label{sec:data_section}
\subsection{MHD wind-driven accretion can reproduce the observed spread of \texorpdfstring{$\log \bigl(M_D/\dot{M}_*\bigr)$}{MD/M*}}
\label{sec:empirical_spread}

To assess whether MHD winds can reproduce both disk lifetimes and the observed spread in $\log \bigl(M_D/\dot{M}_*\bigr)$, we conducted the following analysis. 

Up to now, the spread in $t_{\mathrm{acc},0}$ was considered as a free parameter. As shown by \cite{Tabone2022b} - though a different ansatz for the distribution of $t_{\mathrm{acc},0}$ - the observed disk fraction sets both $\sigma_{t_{\mathrm{acc},0}}$ and the median values of $t_{\mathrm{acc},0}$. Following the work of \cite{Tabone2022b}, we fitted the disk fraction (represented as the fraction of sources that show IR excess) reported in \cite{Fedele_2010} and references therein. We first defined our disk fraction as the ratio of disks that, at a certain time $t$, satisfies the condition
\begin{equation}
    t_{disp} > t.
    \label{eq:tdisp>t}
\end{equation}
Then, we performed a fit using the Complementary Cumulative Distribution Function (from now on, CCDF) assuming a Lognormal distribution of $t_{disp}$.
\begin{figure}[t]
    \centering
    \includegraphics[width=\linewidth]{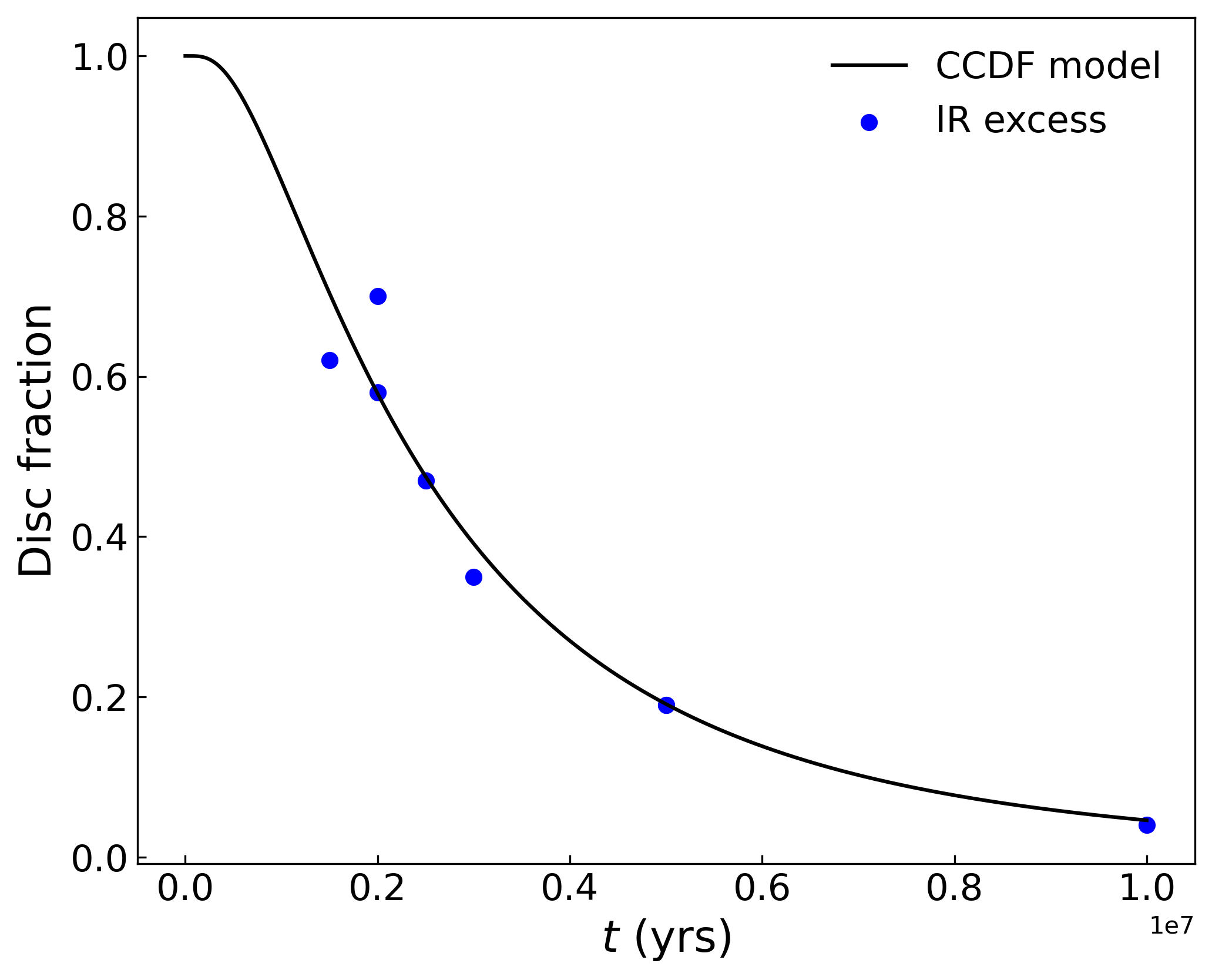}
    \caption{The disk fraction as function of the cluster age (data from \citealt{Fedele_2010} and references therein), with the fit of the CCDF of $t_{disp}$. From the fit, we obtain $t_{disp} = 2.37$ Myrs and $\sigma_{t_{disp}} = 0.37$ dex.}
    \label{fig:disc_frac}
\end{figure}
In this way, we find the parameters that describe the probability for a disk population to satisfy Eq. \eqref{eq:tdisp>t}.

In Fig. \ref{fig:disc_frac}, we show the results of the fit: we find that $t_{disp}$ follows a Lognormal distribution with a median value $\mu_{t_{disp}} = 2.37$ Myrs and a spread $\sigma_{t_{disp}} = 0.37$ dex. After some simple algebra deriving from Eq. \eqref{eq:t_disp}, we find that the median value of the accretion timescale is $\mu_{t_{\mathrm{acc},0}}= 0.95$ Myrs and the spread is $\sigma_{t_{\mathrm{acc},0}} = 0.37$ dex.
With these values, we evaluated the spread of the accretion timescale using Eq. \eqref{eq:sigma_2} with $t=1.5$ Myrs and $\sigma_t = 0$\footnote{For the purposes of the current work, we chose to assume there is no spread in stellar ages.} dex, obtaining $\sigma \bigl( \log \bigl(M_D / \dot{M}_*\bigr) \bigr) = 1.01 $ dex. 

We compared this value with the spread of the accretion timescale evaluated from the data found in \cite{Testi_2022} for the four populations of L1688, Lupus, ChaI and USco. We removed the upper limits from the sample and considered only the systems for which $M_*>0.15 \>M_{\odot}$, in agreement with \cite{Testi_2022}. For L1688 and Lupus, we obtained values of $\sigma \bigl( \log \bigl(M_D / \dot{M}_*\bigr) \bigr)$ that are well-comparable with the one reported above, more specifically $\sigma \bigl( \log \bigl(M_D / \dot{M}_*\bigr) \bigr)_{L1688} = 0.86$ dex, $\sigma \bigl( \log \bigl(M_D / \dot{M}_*\bigr) \bigr)_{Lupus} = 1.08$ dex. As we moved to the older region of ChaI, we obtained the higher spread $\sigma \bigl( \log \bigl(M_D / \dot{M}_*\bigr) \bigr)_{ChaI} = 1.62$ dex, a value that increases even more for USco, the oldest region of the sample, for which we got $\sigma \bigl( \log \bigl(M_D / \dot{M}_*\bigr) \bigr)_{USco} = 2.02$ dex.
The reason behind these higher spreads is that, to evaluate $\sigma \bigl( \log \bigl(M_D / \dot{M}_*\bigr) \bigr)$ we assumed that the distribution of $\log \bigl(M_D/\dot{M}_* \bigr)$ is Normal. This assumption is valid only for young populations of disks (e.g. L1688, Lupus and at most ChaI): as the population evolves in time, we lose disks, and the distribution of $\log \bigl(M_D/\dot{M}_*\bigr)$ significantly deviates from the starting distribution. 

We hence conclude that the spread predicted using Eq. \eqref{eq:sigma_2} can well reproduce the spread of the accretion timescale for young populations of disks. This result is in agreement with the analysis of \cite{Tabone_2022}, but in this work we assumed a Lognormal distribution to fit for $t_{disp}$, as already performed by \cite{Somigliana2023}, and we applied the results to different stellar regions. We furthermore recall that for our calculation we set $\sigma_t = 0$ dex, and we neglected accretion variability and mass uncertainty, so the value of $\sigma \bigl( \log \bigl(M_D / \dot{M}_*\bigr) \bigr)$ that we predicted must be considered as a lower limit.

\subsection{Empirical constraints on \texorpdfstring{$\sigma_{M_0}$}{sigmaM0} and \texorpdfstring{$\phi$}{phi}}
\label{sec:degeneracy}
We conclude this work showing the existence of a degeneracy between $\phi$ and $\sigma_{M_0}$. As pointed out in Sect. \ref{subsec:spread_slope}, the slope of the $M_D - \dot{M}_*$ correlation is affected by both $\phi$ and $\sigma_{M_0}$. From Monte-Carlo simulations it is possible to investigate the $\phi - \sigma_{M_0}$ space in order to constrain for which values of $\phi$ and $\sigma_{M_0}$ it is possible to retrieve a slope $\gamma \in [0.5,0.9]$, with a spread on the $M_D-\dot{M}_*$ correlation $\sigma \in [0.55, 0.71]$ dex (\citealt{Manara_2016}). To do this, we simply evaluated $\gamma$ and the spread of the $M_D-\dot{M}_*$ correlation with a linear regression of the disks residing in the $M_D - \dot{M}_*$ plane. Each Monte-Carlo simulation consists in $10^4$ disks evolving according to Eqs. \eqref{eq:M_D}, \eqref{eq:M_dot_star}, where $M_0 \propto t_{\mathrm{acc},0}^{\phi}$.

\begin{figure}[t]
    \centering
    \includegraphics[width=\linewidth]{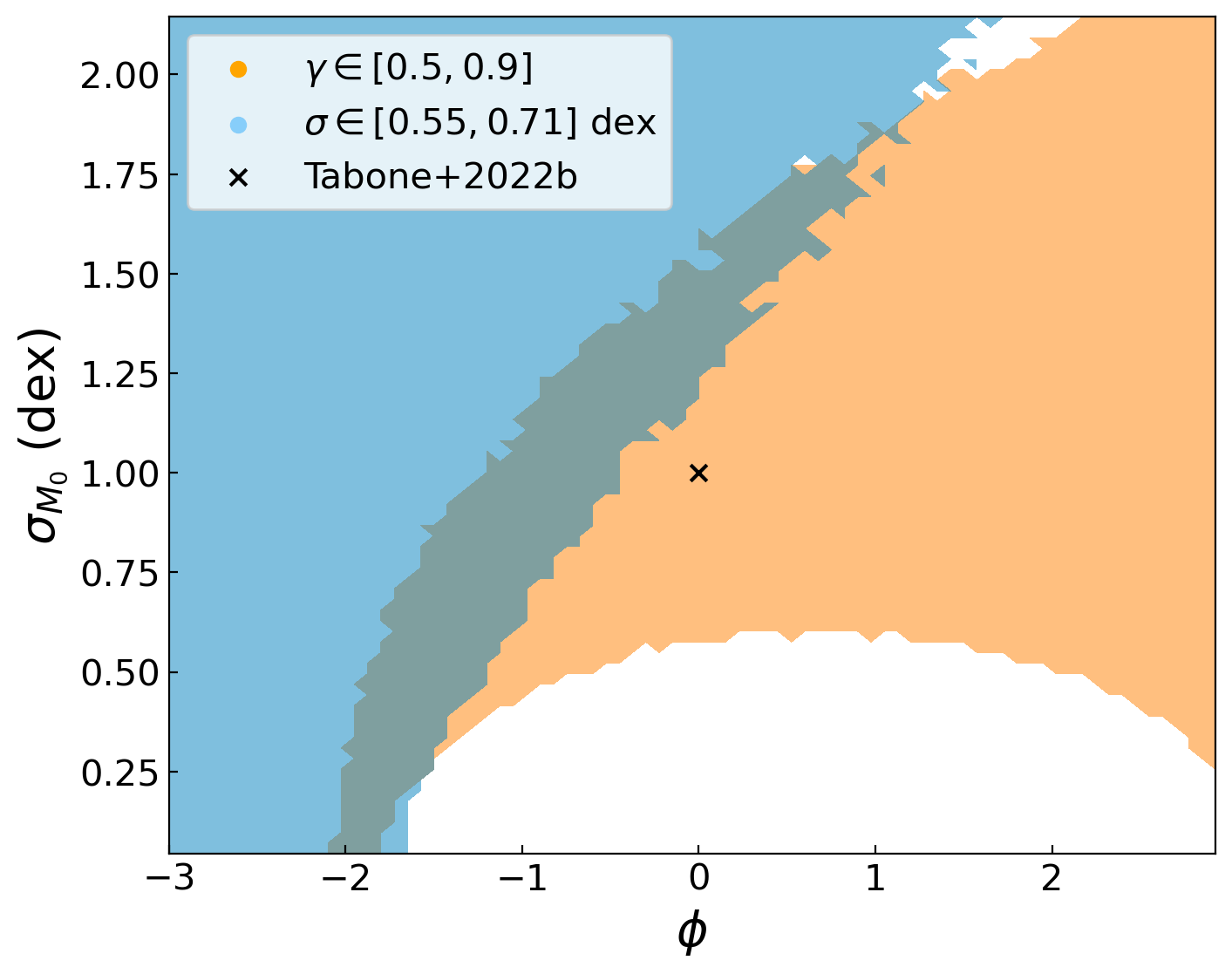}
    \caption{The $\phi - \sigma_{M_0}$ plane with the values of the slope $\gamma$ and the spread of the $M_D-\dot{M}_*$ correlation $\sigma$: we put in evidence in orange the values $\gamma \in [0.5, 0.9]$, while in light blue the values $\sigma \in [0.55, 0.71]$ dex, as they confine the region of values compatible with the values found in \cite{Manara_2016}. The colored regions show the value of $\gamma$ and $\sigma$ for Monte-Carlo disk populations obtained with the corresponding specific values of $\phi$ and $\sigma_{M_0}$. Every Monte-Carlo consists in $10^4$ disks evolving according to Eqs. \eqref{eq:M_D}, \eqref{eq:M_dot_star}, where $M_0 \propto t_{\mathrm{acc},0}^{\phi}$, $\omega = 0.8$, $t=2$ Myrs, and $t_{\mathrm{acc},0}$ follows a Lognormal distribution centered on the values retrieved from the fit in Sect. \ref{sec:empirical_spread}. The cross shows the case investigated by \cite{Tabone2022b}}.
    \label{fig:degeneracy}
\end{figure}

In Fig. \ref{fig:degeneracy}, we show the $\phi - \sigma_{M_0}$ plane with the corresponding values of $\gamma$. The coloured regions in the $\phi - \sigma_{M_0}$ plane show the value of the slope and of its spread obtained via linear regression of a whole population of disks generated with specific values of $\phi$ and $\sigma_{M_0}$. In a similar fashion to the analysis performed in \cite{Lodato2017}, we show in orange the region $\gamma \in [0.5, 0.9]$, while in light blue the region $\sigma \in [0.55, 0.71]$ dex. In this case, $t = 2$ Myrs, $\omega = 0.8$, and $t_{\mathrm{acc},0}$ follows a Lognormal distribution centered on the values derived in Sect. \ref{sec:empirical_spread}. For comparison, we marked with a cross the region covered in \cite{Tabone2022b}.

Following Sect. \ref{subsec:spread_slope}, we found that $\gamma$ tends toward unity as $\sigma_{M_0}$ increases, regardless of the value of $\phi$. However, there is a whole region of the $\sigma_{M_0}-\phi$ plane that allows values of $\gamma \neq 1$: this confirms the existence of a degeneracy between $\phi$ and $\sigma_{M_0}$. Moreover, following Fig. \ref{fig:degeneracy}, we can constrain the ensemble of values of $\sigma_{M_0}$ that allow to have the observed values of $\gamma$ and $\sigma$: we see that only values of $\phi \in [-2, 1]$ and $\sigma_{M_0} \in [0, 1.75]$ dex are allowed. Looking at Fig. \ref{fig:degeneracy}, we understand that for higher values of $\phi$ we need higher values of $\sigma_{M_0}$ in order to retrieve the observed $M_D-\dot{M}_*$ correlation. This can be explained as follows: the stronger the correlation between $M_0$ and $t_{\mathrm{acc,0}}$ is, the bigger the spread in $M_0$ must be in order to break the bent shape shown in Fig. \ref{fig:t_tdisp_sim_1}.

To conclude, we report that the range of values we found for $\sigma_{M_0}$ is in agreement with the results found in \cite{Tabone_2022} for modeling the Lupus population, although there they explored only the case with no correlation between $M_0$ and $t_{\mathrm{acc},0}$. Moreover, we stress that in this section we have shown a procedure to constrain the parameters that rule the evolution of MHD wind-driven disks in the presence of a correlation between $M_0$ and $t_{\mathrm{acc,0}}$.

From Fig. \ref{fig:degeneracy}, we understand that we can obtain values of the slope of the $M_D-\dot{M}_*$ correlation which are compatible with \cite{Manara_2016}, regardless of the initial conditions of the disk populations. We hence conclude that it is possible to derive the slope of the $M_D-\dot{M}_*$ correlation in the MHD wind-driven scenario under a broad range of conditions, a result that shows that the $M_D-\dot{M}_*$ correlation can be easily obtained also for this evolutionary paradigm.

\section{Conclusions}
\label{sec:conclusions}
In this paper we investigated under which conditions the $M_D-\dot{M}_*$ correlation emerges in the MHD wind-driven evolutionary scenario.

We derived and analyzed the parameters that rule the appearing and tuning of the $M_D-\dot{M}_*$ correlation starting from the analytical solutions given by \cite{Tabone2022b}. From our analysis we reach the following conclusions: 

\begin{enumerate}[label=(\roman*)]
    \item We show that introducing a spread $\sigma_{t_{\mathrm{acc},0}}$, we obtain the correlation between the mass of the disk $M_D$ and the accretion onto the central star $\dot{M}_*$. This holds as long as $\sigma_{t_{\mathrm{acc},0}}<\sigma_{M_0}$, a conservative criterion that is compatible with the general ``rule of thumb'' $R^2 \gtrsim 0.5$. In the opposite case, the $M_D-\dot{M}_*$ correlation tends to the "boomerang-shaped" isochrone introduced by \cite{Lodato2017} for the viscous case and studied by \cite{Tabone_2022}, \cite{Tabone2022b} for the wind case, which look significantly different from the observed correlation.

    \item We showed that it is possible to derive a slope and a spread of the $M_D-\dot{M}_*$ correlation comparable to what found in \cite{Manara_2016} varying $\sigma_{M_0}$ and $\sigma_{t_{\mathrm{acc,0}}}$. Hence, it is possible to derive the observed slope and spread of the $M_D-\dot{M}_*$ correlation under a range of initial conditions.

    \item Neglecting the time evolution of the accretion timescale, we derived an analytical expression for the spread of the accretion timescale $\log \bigl(M_D/\dot{M}_*\bigr)$. The predictions arising from this formula are in agreement with the observed order of magnitude of $\sigma(\log \bigl(M_D/\dot{M}_*\bigr) \bigr)$ evaluated for young populations of disks (\citealt{Testi_2022}).

    \item We introduced a correlation $M_0\propto t_{\mathrm{acc},0}^{\phi}$ to understand if we can still derive the observed slope and spread of the $M_D-\dot{M}_*$ correlation. We analytically derived a slope $\gamma=1-1/\phi$ in the initial conditions, and a slope $\gamma=1-\omega$ close to dispersal. 
    
    Furthermore, we noticed that the addition of a minimal spread in the initial disk mass distribution ($\sigma_{M_0}$) affects the slope of the $M_D-\dot{M}_*$ correlation. In particular, we showed that we can fit the $M_D-\dot{M}_*$ correlation with a single power-law because of the existence of $\sigma_{M_0}$.

    \item We studied the degeneracy between $\phi$ and $\sigma_{M_0}$ and we confirmed that we can get values of the slope and of the spread of the $M_D-\dot{M}_*$ correlation in agreement with \cite{Manara_2016} for a broad range of values on both $\sigma_{M_0}$ and $\phi$.
\end{enumerate}

This work is the natural follow-up of the work of \cite{Tabone_2022} and \cite{Tabone2022b}. We investigated in more detail the emergence of the $M_D-\dot{M}_*$ correlation and demonstrated that the slope of the correlation is controlled by the correlation among the parameters describing the initial conditions (physically speaking, the star formation process). With this, we reached the fundamental conclusion that MHD wind-driven disks can reproduce the observed slope of the $M_D-\dot{M}_*$ correlation under a broad range of initial conditions. Therefore, no fine tuning of the model parameters is needed in the MHD scenario, but reproducing the correlation does exclude some regions of the parameter space, as shown by Figs. \ref{fig:sigmaM0_sigmatacc0}, \ref{fig:degeneracy}. While in this paper we focused only on the MHD wind scenario, our work implies that the observed $M_D-\dot{M}_*$ correlation cannot be used to discriminate between MHD winds and viscous model. However, the correlation still constrains the model parameters for each scenario, as shown, for example, by our work and \cite{Lodato2017}.

\begin{acknowledgements}
We are thankful to the anonymous referee who helped improve the clarity of the paper.
The authors acknowledge support from the European Union (ERC Starting Grant DiscEvol, project number 101039651), from Fondazione Cariplo, grant No. 2022-1217, from the European Union’s Horizon 2020 research and innovation programme under the Marie Sklodowska-Curie grant agreement No 823823 (Dustbusters RISE project), and from the ERC Synergy Grant “ECOGAL” (project ID 855130). Views and opinions expressed are, however, those of the authors only and do not necessarily reflect those of the European Union or the European Research Council. Neither the European Union nor the granting authority can be held responsible for them. We thank Carlo Manara for the very useful comments he provided for ensuring the correctness of this work. Moreover, we thank 
Giacomo Lucertini for the useful discussions in statistics, Edoardo Merli for IT support, Davide Giovagnoli for interesting dialogues in mathematics, Enrico Ragusa and Chiara Scardoni for the many chats in physics.
\end{acknowledgements}

\section*{Data availability}
All data used in this work are publicly available and can be found at the corresponding cited works. The scripts that have been used for this work will be shared under reasonable request to the corresponding author.

\bibliographystyle{aa}
\bibliography{main}

\begin{appendix}
\section{The \texorpdfstring{$\sigma_{M_0} - \sigma_{t_{\mathrm{acc,0}}}-R^2$}{M0-tacc,0-R2} plane for different values of \texorpdfstring{$\omega$}{omega}}
\label{appendix:sigma_M0-sigma_tacc0}
\begin{figure*}
    \sidecaption
    \centering
    \includegraphics[width=0.6\linewidth]{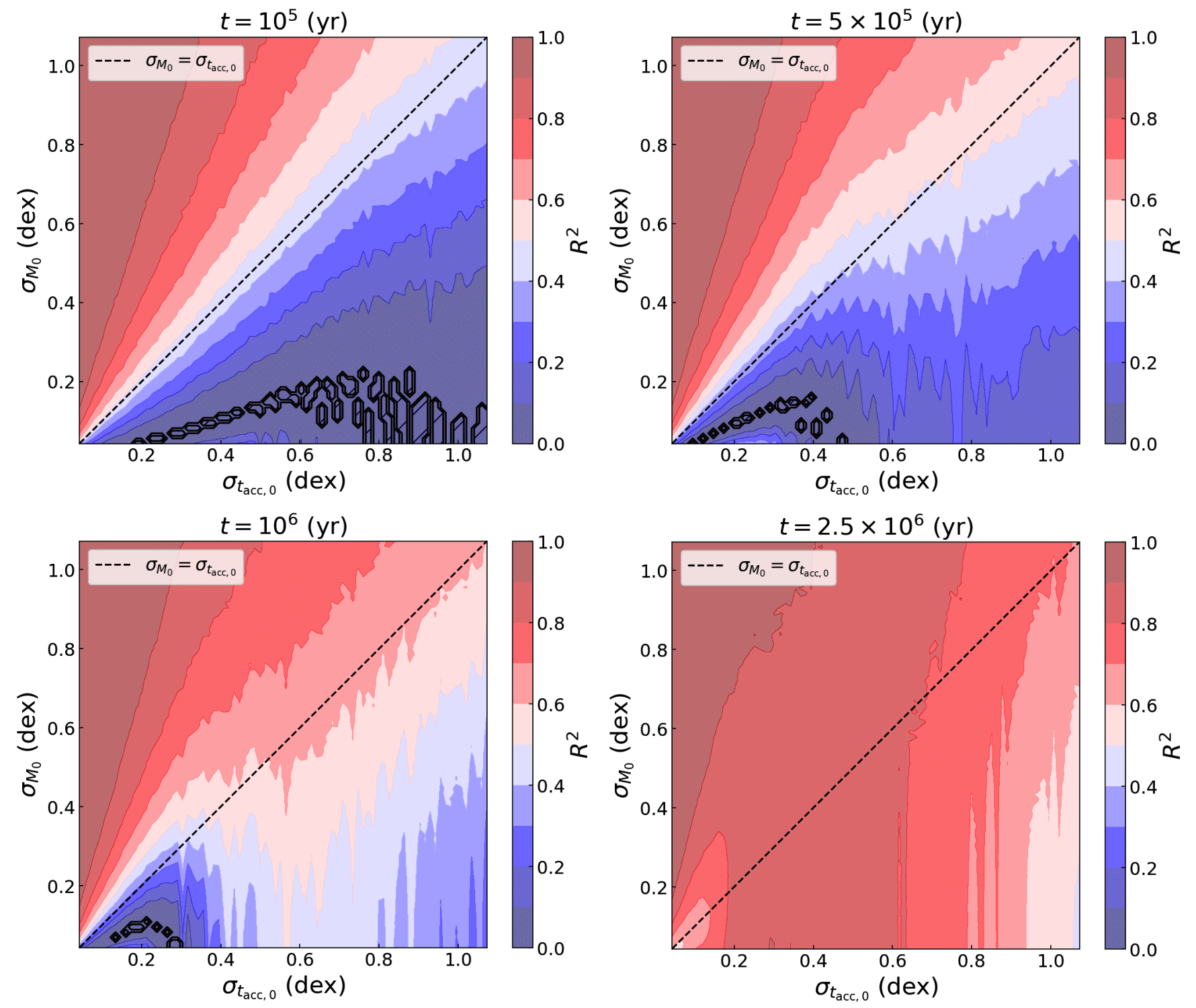}
    \caption{The $\sigma_{M_0}-\sigma_{t_{\mathrm{acc,0}}}$ plane with the values of $R^2$ for different values of age $t$. The colored dots in every plot are Monte-Carlo simulations of $10^5$ proto-planetary disks with $\omega=0.2$, evolving following Eqs. \eqref{eq:M_D}, \eqref{eq:M_dot_star}, where $M_0$ and $t_{\mathrm{acc,0}}$ follow a Lognormal distribution centered on the natural logarithm of $0.1\>M_{\odot}$ and $1$ Myrs, respectively. The hatched contours represent the regions where there is no correlation between $M_D $ and $\dot{M}_*$, while the dashed line represents the points where $\sigma_{M_0} = \sigma_{t_{\mathrm{acc,0}}}$.} 
    \label{fig:sigmaM0-sigmatacc0_omega=0.2}
\end{figure*}

\begin{figure*}
    \sidecaption
    \centering
    \includegraphics[width=0.6\linewidth]{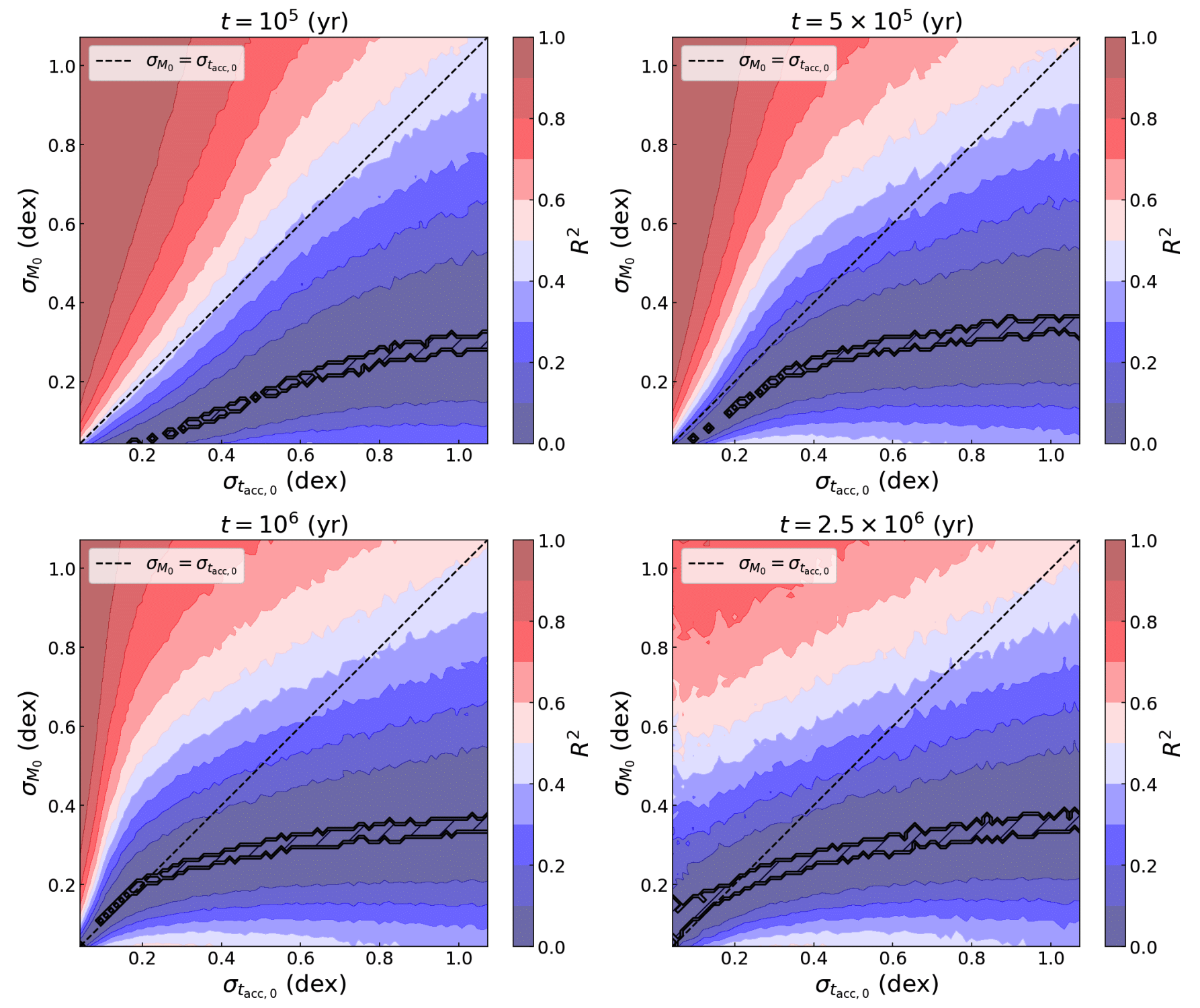}
    \caption{The $\sigma_{M_0}-\sigma_{t_{\mathrm{acc,0}}}$ plane with the values of $R^2$ for different values of age $t$. The colored dots in every plot are Monte-Carlo simulations of $10^5$ proto-planetary disks with $\omega=1.0$, evolving following Eqs. \eqref{eq:M_D}, \eqref{eq:M_dot_star}, where $M_0$ and $t_{\mathrm{acc,0}}$ follow a Lognormal distribution centered on the natural logarithm of $0.1\>M_{\odot}$ and $1$ Myrs, respectively. The hatched contours represent the regions where there is no correlation between $M_D $ and $\dot{M}_*$, while the dashed line represents the points where $\sigma_{M_0} = \sigma_{t_{\mathrm{acc,0}}}$.} 
    \label{fig:sigmaM0-sigmatacc0_omega=1.0}
\end{figure*}

\begin{figure}
    \centering
    \includegraphics[width=\linewidth]{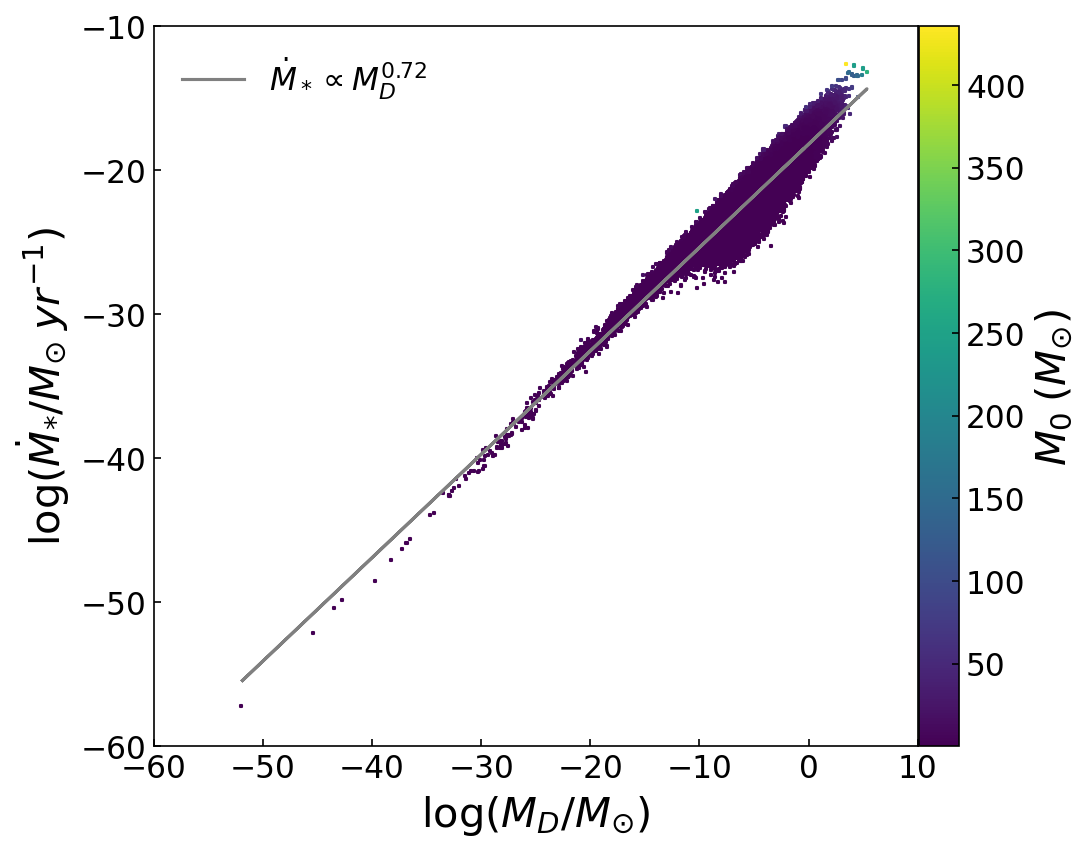}
    \caption{The $M_D - \dot{M}_*$ correlation with $\omega = 0.2$ and $t=2.5$ Myrs. The colored dots are Monte-Carlo simulations of $10^5$ proto-planetary disks evolving following Eqs. \eqref{eq:M_D}, \eqref{eq:M_dot_star} where $M_0$ and $t_{\mathrm{acc},0}$ follow a Lognormal distribution centered, respectively, on the natural logarithm of $0.1 \> M_{\odot}$ with a spread of 0.8 dex and on the natural logarithm of $1$ Myrs with a spread of 0.4 dex. The gray line represents the best fit to the population, resulting in a slope $\gamma=0.72$.}
    \label{fig:No_Corr_omega=0.2.png}
\end{figure}

In Figs. \ref{fig:sigmaM0-sigmatacc0_omega=0.2}, \ref{fig:sigmaM0-sigmatacc0_omega=1.0} we show the $\sigma_{M_0}-\sigma_{t_{\mathrm{acc,0}}}$ plane for $\omega=0.2$ and $\omega=1.0$, respectively. All the other parameters are the same as the ones presented in Fig. \ref{fig:sigmaM0-sigmatacc0_omega=0.8}. We highlight that the ``rule of thumb'' derived in Sect. \ref{sec:no_corr} holds also for these ``extreme'' values of $\omega$. Moreover, we note that, in Fig. \ref{fig:sigmaM0-sigmatacc0_omega=0.2} for evolved populations our general criterion $R^2 \gtrsim 0.5$ does not correspond anymore necessarily to the region $\sigma_{M_0}>\sigma_{t_{\mathrm{acc,0}}}$. The reason for this is shown in Fig. \ref{fig:No_Corr_omega=0.2.png}: for evolved populations, when $\omega \to 0$, the disks collect around the evolved region of the "boomerang-shaped" isochrone, thus being effectively fitted with a single power-law. However, we stress that our criterion is very general, therefore any simulated population based on Figs. \ref{fig:sigmaM0-sigmatacc0_omega=0.2}, \ref{fig:sigmaM0-sigmatacc0_omega=1.0}, \ref{fig:sigmaM0-sigmatacc0_omega=0.8}, must be checked directly in the $M_D-\dot{M}_*$ plane.

\section{The slope of the \texorpdfstring{$M_D - \dot{M}_*$}{MD-M*} correlation when \texorpdfstring{$t_{\mathrm{acc},0}$}{tacc0} is constant: analytics}
\label{appendix:tacc0_const}

Considering $t_{\mathrm{acc},0}$ as a constant value, Eq. \eqref{eq:M_dot_star} can be seen as 
\begin{equation}
\label{eq:M_dot_star_mesh}
\dot{M}_*= \frac{M_D}{2t_{\mathrm{acc},0}(1+f_{M,0})} \Biggl(1-\frac{\omega t}{2 t_{\mathrm{acc},0}} \Biggr)^{-1}.
\end{equation}
Taking the hint from the work of \citealt{Manara_2016}, it is possibile to investigate the analytical behavior of disks evolving according to Eq. \eqref{eq:M_dot_star_mesh}, assuming that 
\begin{equation}
\label{eq:M_dot_prop}
\dot{M}_* \propto M_D^{\gamma}.
\end{equation}
Performing the logarithmic derivative on Eq. \eqref{eq:M_dot_prop}, the slope $\gamma$ results in the constant 
\begin{equation}
\label{eq:gamma}
\gamma = \frac{M_D}{\dot{M}_*} \frac{\mathrm{d}\dot{M}_*}{\mathrm{d}M_D} =1.
\end{equation}

\section{The spread of the \texorpdfstring{$M_D-\dot{M}_*$}{MD-M*} correlation as function of the spread in \texorpdfstring{$t$}{t} and \texorpdfstring{$t_{\mathrm{acc}}$}{tacc}: analytics}
\label{appendix:spread}
Starting from Eqs. \eqref{eq:M_D}, \eqref{eq:M_dot_star}, the ratio $M_D / \dot{M}_*$ reads
\begin{equation}
\frac{M_D}{\dot{M}_*} = (1+f_{M,0})\>(2t_{\mathrm{acc},0}-\omega t).
\end{equation}
Neglecting the covariance term between $t$ and $t_{\mathrm{acc},0}$, the spread of $\log \bigl(M_D/\dot{M}_*\bigr)$ correlation is obtained with a simple propagation of errors as
\begin{equation}
\begin{split}
\sigma \bigl( \log \bigl(M_D/\dot{M}_*\bigr) \bigr) & = \Biggl[ \sigma_{t_{\mathrm{acc},0}}^2 \Biggl( \frac{\partial \bigl( \log \bigl(M_D/\dot{M}_*\bigr) \bigr)}{\partial \log \bigl(t_{\mathrm{acc},0} \bigr)} \Biggr)^2 +\\
&  \sigma_{t}^2 \Biggl( \frac{\partial \bigl(\log \bigl(M_D/\dot{M}_* \bigr) \bigr)}{\partial \log\bigl(t\bigr)} \Biggr)^2  \Biggr]^{1/2}, 
\end{split}
\label{eq:sigma_1}
\end{equation}
which gives Eq. \eqref{eq:sigma_2} after some simple algebra. We stress that the quantity $\sigma_{t_{\mathrm{acc},0}}$ is the spread of the accretion timescale. This quantity, due to dispersal of the fastest evolving disks, decreases as a function of time. However, for our calculations, we keep it fixed to its initial value.

\section{The slope of the \texorpdfstring{$M_D-\dot{M}_*$}{MD-M*} correlation when \texorpdfstring{$M_0 \propto t_{\mathrm{acc},0}^{\phi}$}{M0=tacc0phi}: analytics}
\subsection{The slope for \texorpdfstring{$t/t_{disp} <<1$}{t/tdisp<<1}}
\label{appendix:t/tdisp<<1}
Injecting Eq. \eqref{eq:M_0relt_0} into Eq. \eqref{eq:M_D}, we find that, in the initial conditions of the evolution of the disk, namely when $t/t_{disp} << 1$, 
\begin{equation}
t_{\mathrm{acc},0} \propto M_D^{1/{\phi}}.
\end{equation}
Including this result in Eq. \eqref{eq:M_dot_star}, after the trivial substitution of Eq. \eqref{eq:M_0relt_0} in Eq. \eqref{eq:M_dot_star}, we obtain 
\begin{equation}
\label{eq:M_dot_initial}
\dot{M}_* \propto M_D^{1-1/\phi}.
\end{equation}
The logarithmic derivative gives the slope in the form 
\begin{equation}
\gamma = \frac{M_D}{\dot{M}_*} \frac{\mathrm{d}\dot{M}_*}{\mathrm{d}M_D} = 1-\frac{1}{\phi}.
\label{eq:gamma_alpha}
\end{equation}

\subsection{The slope for \texorpdfstring{$t \sim t_{disp}$}{t=tdisp}}
\label{appendix:t=tdisp}
We derive the slope in the final life-time conditions of a disk starting by choosing $\epsilon$ to be arbitrarily small. Then, inserting 
\begin{equation}
\frac{t}{t_{disp}} = 1- \epsilon
\end{equation}
in Eqs. \eqref{eq:M_D}, \eqref{eq:M_dot_star} after the previous substitution of Eq. \eqref{eq:M_0relt_0}, $\epsilon$ is given by
\begin{equation}
\epsilon \propto \Biggl( \frac{M_D}{t_{\mathrm{acc},0}^{\phi}} \Biggr)^{\omega}.
\end{equation}
After some algebra, it is obtained that 
\begin{equation}
\label{eq:M_dot_tdisp}
\dot{M}_* \propto t_{\mathrm{acc},0}^{\>\phi  \omega -1} \> M_D^{1-\omega},
\end{equation}
which leads to a slope of the form 
\begin{equation}
\gamma = \frac{M_D}{\dot{M}_*} \frac{\mathrm{d}\dot{M}_*}{\mathrm{d}M_D} = 1-\omega.
\end{equation}

\section{The spread of the \texorpdfstring{$M_D - \dot{M}_*$}{MD-M*} correlation when \texorpdfstring{$M_0 \propto t_{\mathrm{acc},0}^{\phi}$}{M0=tacc0phi}}
\label{appendix:spread_propto}
\subsection{The spread when $M_0 \propto t_{\mathrm{acc},0}^{\phi}$ and $t << t_{disp}$}
Following Sect. \ref{sec:spread}, the calculations for the spread when $M_0 \propto t_{\mathrm{acc},0}^{\phi}$ lead to 
\begin{equation}
\sigma \bigl( \log \bigl(M_D/\dot{M}_*\bigr) \bigr) = \sigma_{t_{\mathrm{acc}}}.
\end{equation}
This shows the trivial fact that, in the initial conditions, the spread of the correlation depends uniquely on the spread of $t_{\mathrm{acc},0}$, a result already shown in Fig. \ref{fig:Std_dev_sigma_t=0}.
\begin{figure}[t]
    \centering   \includegraphics[width=\linewidth]{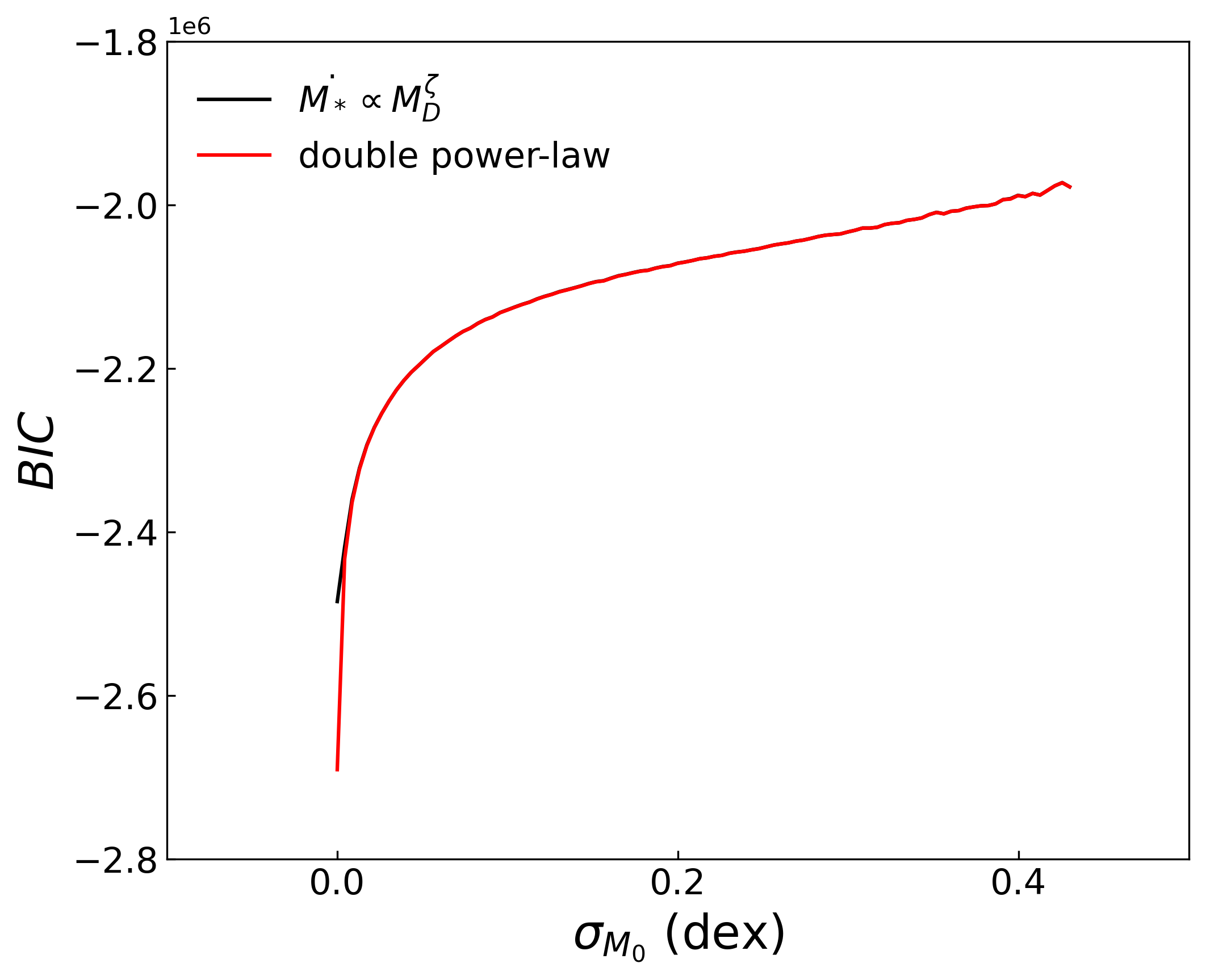}
    \caption{The Bayesian Information Criterion $BIC$ as function of $\sigma_{M_0}$ in a range $\sigma_{M_0} \> \in \> [0, 0.43]$ dex. The $BIC$ is obtained for Eq. \eqref{eq:double_powerlaw} (in red) and Eq. \eqref{eq:single_powerlaw} (in black) from the fit to the $\dot{M}_* -M_D$ plane obtained for different values of $\sigma_{M_0}$, while the simulated disk populations have the same set of parameters described in Fig. \ref{fig:t_tdisp_sim_1}.}
    \label{fig:BIC}
\end{figure}

\subsection{The spread when $M_0 \propto t_{\mathrm{acc},0}^{\phi}$ and $t \sim t_{disp}$}
A similar result is obtained when $t \sim t_{disp}$, obtaining
\begin{equation}
\label{eq:sigma_t_sim_tdisp}
\sigma \bigl( \log \bigl(M_D/\dot{M}_* \bigr) \bigr) = \epsilon \>\sigma_{t_{\mathrm{acc}}}.
\end{equation}
It is clear that the more $t$ approaches the value of $t_{disp}$, the littler is the value of $\epsilon$. In the limit for $\epsilon \to 0$, Eq. \eqref{eq:sigma_t_sim_tdisp} gives $\sigma \bigl( \log \bigl(M_D / \dot{M}_* \bigr) \bigr) \to 0$: a valid result that reflect the fact that when the disk is about to die, the slope of the $M_D - \dot{M}_*$ correlation inevitably tends toward the value $1-\omega$.

\section{The likelihood and the \texorpdfstring{$BIC$}{BIC} in a wider \texorpdfstring{$\sigma_{M_0}$}{sigmaM0} range}
\label{appendix:likelihood_BIC}

In Fig. \ref{fig:BIC} we show the same results of Fig. \ref{fig:BIC_smallspread}, in a range $\sigma_{M_0} \> \in \> [0, 1]$. As can be seen, for significant values of $\sigma_{M_0}$, the $\dot{M}_* - M_D$ loses the double power-law behavior in favor of a single power-law written in the form of Eq. \eqref{eq:single_powerlaw}.
\end{appendix}

\end{document}